\documentclass[usenatbib,fleqn]{mn2e}
\usepackage{hyperref,amsmath,amssymb,times,graphicx,xcolor,fleqn}
\usepackage{etoolbox}
\makeatletter
\patchcmd\@combinedblfloats{\box\@outputbox}{\unvbox\@outputbox}{}{\errmessage{\noexpand patch failed}}
\makeatother

\title[Supermassive black hole binaries]{Evolution of supermassive black hole binaries and tidal disruption rates in nonspherical galactic nuclei}
\author[K. Lezhnin \& E. Vasiliev]{
Kirill Lezhnin$^{1,2}$\thanks{klezhnin@princeton.edu, eugvas@lpi.ru},
Eugene Vasiliev$^{3,4,5}$\\
$^{1}$Department of Astrophysical Sciences, Princeton University, Princeton, New Jersey 08544, USA\\
$^{2}$National Research Nuclear University MEPhI, Kashirskoe sh. 31, Moscow, 115409, Russia\\
$^{3}$Institute of Astronomy, Madingley road, Cambridge, CB3 0HA, UK\\
$^{4}$Rudolf Peierls Centre for Theoretical Physics, 1 Keble road, Oxford, OX1 3NP, UK\\
$^{5}$Lebedev Physical Institute, Leninsky prospekt 53, Moscow, 119991, Russia}

\begin{document}
\date{Accepted 2019 January 11. Received 2018 December 15; in original form 2018 October 20}
\maketitle

\begin{abstract}
Binary supermassive black holes (SMBH) are expected to form naturally during galaxy mergers. After the dynamical friction phase, when the two SMBHs become gravitationally bound to each other, and a brief stage of initial rapid hardening, the orbit gradually continues to shrink due to three-body interactions with stars that enter the loss cone of the binary. Using the stellar-dynamical Monte Carlo code \textsc{Raga}, we explore the co-evolution of the binary SMBH and the nuclear star cluster in this slow stage, for various combinations of parameters (geometry of the star cluster, primary/secondary SMBH mass, initial eccentricity, inclusion of stellar captures/tidal disruptions).
We compare the rates of stellar captures in galactic nuclei containing a binary to those of galaxies with a single SMBH. At early times, the rates are substantially higher in the case of a binary SMBH, but subsequently they drop to lower levels. Only in triaxial systems both the binary hardening rates and the capture rates remain sufficiently high during the entire evolution.
We find that the hardening rate is not influenced by star captures, nor does it depend on eccentricity; however, it is higher when the difference between the black hole masses is greater. 
We confirm that the eccentricity of the binary tends to grow, which may significantly shorten the coalescence time due to earlier onset of gravitational-wave emission.
We also explore the properties of the orbits entering the loss cone, and demonstrate that it remains partially full throughout the evolution in the triaxial case, but significantly depleted in the axisymmetric case. 
Finally, we study the distribution of ejected hypervelocity stars and the corresponding mass deficits in the central parts of the galaxies hosting a binary, and argue that the missing mass is difficult to quantify observationally.
\end{abstract}

\section{Introduction}

It is commonly accepted that supermassive black holes (SMBHs) reside in the centres of most galaxies \citep{KormendyHo2013, Graham2016}. In the conventional cosmological paradigm, larger galaxies are formed through hierarchical mergers of smaller galaxies. Thus the merger remnants may contain multiple SMBHs, which will eventually sink towards the galaxy centre due to dynamical friction, and form a binary (or a higher-multiplicity) system. The subsequent evolution of a binary SMBH is driven by a combination of several effects, which all lead to a gradual shrinking of the binary orbit (\citealt{Begelman1980}, see also more recent reviews by \citealt{MerrittMilos2005}, \citealt{Colpi2014}). Three-body interactions between the pair of SMBHs and stars of the galactic nucleus typically result in an ejection of the lighter of the three bodies (the star), extracting the energy from the binary orbital motion \citep{Quinlan1996}. Torques from the circumbinary gaseous discs also lead to the shrinking of the binary \citep[e.g.,][]{Mayer2013, Rafikov2016, Tang2017}, although this process appears to be effective only when the separation becomes small enough \citep{Lodato2009}. Finally, when the separation between the two SMBHs reaches $\lesssim 10^{-3}$~pc, the emission of gravitational waves (GW) becomes efficient enough and leads to the ultimate coalescence of the SMBHs.

Observational evidence for the existence of binary SMBHs is scarce, and to date only a few candidate objects have been found through direct imaging \citep{Rodriguez2006, Kharb2017}. Among various indirect observational signatures (see \citealt{Komossa2006}, \citealt{Dotti2012} for a review) are tidal disruption events (TDEs), when a star passing too close to one of the SMBH is torn apart by the tidal forces.
The presence of a second SMBH modifies the lightcurve \citep[e.g.][]{Ricarte2016, Coughlin2017, Vigneron2018}, and rates of TDEs in galactic nuclei hosting single and binary SMBHs may also differ substantially. In particular, during the early stage of binary evolution ($\sim 10^5-10^6$~yr after its formation), TDE rates could be dramatically elevated, reaching $10^{-2}-1$ events per year \citep{Ivanov2005, Chen2009, Chen2011, LiuChen2013, Li2017}, while for the remaining part of its evolution the rates could be much lower \citep{Chen2008}. On the other hand, high TDE rates could also result from other factors, such as triaxial geometry of the galactic nucleus \citep{MerrittPoon2004, Vasiliev2014a, LezhninVasiliev2016}, or steep density cusps \citep{StoneVelzen2016, Stone2018}. Finally, a double TDE within a single galaxy (not yet observed), separated by an interval of a few hundred days, would be a possible signature of two stars in a stellar binary being captured by two components of a SMBH binary \citep{Coughlin2018, WuYuan2018}. If TDE by binary SMBHs can be reliably distinguished from those by single SMBHs, they would serve as a valuable probe of the cosmic population of low-mass binary SMBHs in upcoming large-scale surveys \citep{Thorp2018}.

The lifetime of a SMBH binary in a gas-poor galactic nucleus is determined by the rate of its interaction with stars that come to its vicinity: some of them would be tidally disrupted, but most will be eventually ejected with high velocity, carrying away the energy from the binary. Early studies identified a possible bottleneck related to the eventual depletion of the reservoir of interacting stars (so-called final-parsec problem, \citealt{MilosMerritt2001}). However, now it is commonly believed to be an artifact of the assumption of spherical geometry. $N$-body simulations of merger remnants produce less symmetric systems and do not exhibit the stalling of the binary evolution \citep{Preto2011, Khan2011}; simulations of isolated non-spherical systems confirm this conclusion \citep{Berczik2006, Vasiliev2014b}. The disappearance of the final-parsec problem has been linked to the existence of a sufficiently large population of so-called centrophilic orbits (those which may approach the binary closely enough, although not at every pericentre passage), sustaining the binary hardening and leading to its coalescence on a timescale $\lesssim 1$~Gyr \citep{Vasiliev2015b, Gualandris2017}. Fully cosmological simulations cannot explicitly follow the dynamical evolution of SMBH binaries, but can still be used to estimate their population under reasonable assumptions about unresolved physical processes \citep[e.g.,][]{Kelley2017, Ryu2018}; the estimated lifetimes lie in the range from $10^8$ to few${}\times10^9$~yr.

The stellar-dynamical processes in galactic nuclei with single or binary SMBH can be studied with various approaches, but for several reasons discussed later in this paper, we choose the Monte Carlo simulation method, implemented in the code \textsc{Raga} \citep{Vasiliev2015a}. 
It has been used previously to study the TDE rates around single SMBH \citep{Vasiliev2014a, LezhninVasiliev2016} and the evolution of binary SMBH \citep{Vasiliev2015b, Vasiliev2016}, but not simultaneously. In the present study, we extend it to include both processes, and conduct a large suite of simulations, exploring the influence of several properties of the system: the total mass of the binary SMBH, its mass ratio, initial eccentricity of its orbit, the geometry of the stellar distribution (spherical, axisymmetric or triaxial), its mean rotation, and the inclusion or neglect of stellar captures. 
We confirm the earlier conclusions of \citet{Vasiliev2015b} and \citet{Gualandris2017} that the final-parsec problem can be avoided only in triaxial systems, and that the hardening rate of the binary due to interactions with stars is almost independent of its eccentricity, but is higher for binaries with a larger mass ratio.
We also find that the occasional captures of stars do not have any influence on the hardening rate. The difference between the capture rates in galactic nuclei hosting a binary SMBH or a single SMBH, but identical other properties, depends on the evolutionary stage and the geometry of the stellar distribution. At early times, the capture rates by binary SMBHs are much higher than by single SMBHs, but subsequently they drop to lower values. In spherical or axisymmetric systems they remain very low throughout the evolution, in agreement with earlier studies \citep{Chen2008}, but in triaxial systems they may be comparable to or only moderately lower than the single SMBH capture rates.
We examine the properties of stellar orbits before and after interaction with the binary. We demonstrate that the loss cone remains well populated in triaxial systems, but significantly depleted otherwise. Hypervelocity stars ejected from the galaxy by the slingshot interaction have a roughly exponential distribution in velocity, reaching few thousand km/s.
Finally, we study the impact of the binary on the structure of the galactic nucleus, in particular, the erosion of the stellar cusp. We argue that while it definitely takes place, it is very hard to measure observationally the amount of missing mass or to relate it to the properties of the binary.

The paper is organized as follows. We start by summarizing the classical loss-cone theory and its extension to non-spherical galactic nuclei in Section~\ref{sec:losscone}, referring the reader to Chapter~6 in \citet{MerrittBook} for a more detailed treatment. Then in Section~\ref{sec:methods} we review various approaches used for studying the evolution of binary SMBHs and TDE rates, and justify our choice of the Monte Carlo method. Section~\ref{sec:ic} presents our initial conditions for single and binary SMBH simulations.
The results of Monte Carlo simulations and the various aspects of the evolution of galactic nuclei are discussed in Section~\ref{sec:results}, and finally Section~\ref{sec:summary} summarizes our findings.

\section{Loss-cone theory for single and binary SMBH}  \label{sec:losscone}

A star of mass $M_\star$ and radius $R_\star$ is tidally disrupted by the black hole of mass $M_\bullet$ if its distance of closest approach is less than
\begin{subequations}  \label{eq:rcapt}
\begin{align}
r_t \equiv \big(\eta^2\,M_\bullet / M_\star \big)^{1/3} R_\star,
\end{align}
where $\eta\sim 1$ is a dimensionless factor depending on the stellar structure. If the black hole is sufficiently massive, it may swallow stars entirely. The equivalent critical distance (computed using a general-relativistic capture criterion, but expressed in terms of a non-relativistic Keplerian orbit) for a non-spinning SMBH is
\begin{align}
r_c \equiv 8\,G\,M_\bullet / c^2.
\end{align}
\end{subequations}
The largest of the two values is called the loss-cone radius $r_\mathrm{LC}$. A more rigorous treatment \citep[e.g.][Figure 1]{ServinKesden2017} shows that this simple approximation only slightly underestimates the exact relativistic solution (the largest error $\sim 20\%$ is at $r_t=r_c$, which for $\eta=1$ corresponds to $M_\bullet \simeq 1.5\times 10^7\,M_\odot$), and that  stars can still be disrupted (and not captured) by SMBH up to $\sim 4.5$ times more massive than this critical value. 
It is also common to formulate the loss-cone boundary in terms of the critical angular momentum $L_\mathrm{LC}\equiv\sqrt{2\,G\,M_\bullet\,r_\mathrm{LC}}$.

The concept of loss cone may be extended to the case of a binary SMBH \citep[e.g.,][]{Yu2002,MilosMerritt2003}, describing the distance of closest approach to the binary centre-of-mass at which the incoming star undergoes a complex three-body scattering process and is ultimately ejected (if it was not captured by either of the two SMBHs in this process). Naturally, the critical distance is of order $a$ -- the semimajor axis of the binary, which is much larger than the loss-cone radius of an isolated black hole.

Regardless of the physical nature of the loss cone, the stars on the orbits with pericentre distances smaller than the critical distance are essentially lost within one dynamical time (either destroyed or ejected, in the latter case they may still return back at some later time, but this may well be viewed as an incomplete scattering event). The important question is whether the timescale of repopulation of these loss-cone orbits is longer or shorter than the dynamical time. In the former case (empty loss cone), the rate of interactions between the stars and the SMBH(s) is supply-limited and depends on the mechanism of repopulation, while in the latter case (full loss cone), this repopulation is efficient enough that the rate no longer depends on it, and is simply proportional to the geometrical size of the loss cone.

In the (idealized) case of a spherically-symmetric stellar system and the SMBH sitting in its centre, only the orbits with low enough angular momentum can enter the loss cone. Since in this case the angular momentum of a star can only be modified by two-body (collisional) relaxation, which is very inefficient in galactic nuclei with large enough number of stars ($\gg 10^6$), the loss cone is typically quite empty. On the other hand, in non-spherical systems the angular momentum changes due to the torques from the global stellar potential (and not because of individual encounters), i.e., through collisionless processes. Even in this case, the loss cone needs not be full, but is typically at least partially refilled, so that the rate of interactions is a significant fraction of the `limiting' full-loss-cone rate. 
A clear signature of this regime is that a star can enter the loss cone with any value of the angular momentum smaller than $L_\mathrm{LC}$. By contrast, in the opposite (empty loss cone) regime, the typical change of angular momentum per one orbital period is small, hence a star is likely to enter the loss cone with the angular momentum only slightly smaller than the critical value $L_\mathrm{LC}$. By studying the angular-momentum distribution of stars entering the loss cone, one may draw conclusions about the dominant mechanism of the loss-cone repopulation (collisional or collisionless).
Collisional processes are believed to be unimportant in all but the densest galactic nuclei.

\section{Methods}  \label{sec:methods}

The dynamical evolution of a stellar system surrounding a single or a binary SMBH may be studied using various approaches. 
$N$-body simulations offer the most straightforward way of incorporating all relevant physical processes (gravitational encounters of stars with the components of the binary and between stars themselves, stellar evolution, gas dynamics, etc.), but they are very computationally expensive. 
Broadly speaking, existing simulation methods fall into two categories (see \citealt{DehnenRead2011} for a review).

`Collisional' codes accurately compute the gravitational force, typically using the direct-summation approach (with the cost proportional to $N^2$), and employ high-accuracy time integration schemes, often combined with regularization techniques to handle close encounters between particles, or tightly bound subsystems. Because of this, they cannot presently be used for systems with the number of particles significantly in excess of $10^6$, even when running on special-purpose hardware \citep{Makino1996} or modern GPU-accelerated parallel supercomputers \citep[e.g.][]{Wang2015, Li2017}%
\footnote{\citet{Khan2016} performed a $N\simeq 6\times10^6$ simulation of a high-redshift galaxy merger, using a direct-summation code; however, the evolution lasted only $\sim 10^7$~yr before the GW-induced coalescence of the binary SMBH.}. 
Real galaxies, of course, contain a much larger number of stars, so any simulation with a lower number of particles must be appropriately scaled, which is tricky when the various physical processes have different dependence on $N$. In particular, the two-body relaxation rate is $\propto N^{-1}\,\log N$, being significantly higher in scaled-down $N$-body systems than in real galaxies. This distorts the interplay between relaxation-driven and collisionless repopulation of the loss cone and makes it difficult to draw firm conclusions about the long-term evolution of real galaxies \citep{Vasiliev2015b}.

On the other hand, `collisionless' codes sacrifice the accuracy of force computation and time integration to allow a significantly larger number of particles, reducing (but not eliminating) the impact of two-body relaxation. The gravitational force is computed approximately, using grid, tree-code or fast-multipole methods, which scale as $\mathcal{O}(N\,\log N)$ or even $\mathcal{O}(N)$; the force is also softened at small distances, which additionally suppresses the two-body relaxation (but only moderately). We stress that these methods are called `collisionless' only because they \textit{cannot} properly handle two-body encounters (collisions), not because they somehow completely mitigate their influence: at a fixed number of particles, the two-body relaxation rate in these approaches is only a factor of few lower than in collisional codes \citep{HernquistBarnes1990}. It is only the use of a substantially larger $N$ that allows them to be nearly relaxation-free. However, in their standard form these methods are unsuitable for studying close encounters between stars and SMBH(s) by design.

It is not impossible to imagine a hybrid method that would combine a large number of particles and hence negligible two-body relaxation with an accurate treatment of close encounters between the particles and SMBHs. Indeed, the fast-multipole method can achieve the accuracy of force computation comparable to the direct-summation approach, while still retaining its $\mathcal{O}(N)$ scaling \citep{Dehnen2014}. \citet{Gualandris2017} used this method with up to $N\sim10^8$ particles to study the rate of loss-cone repopulation in non-spherical galaxies, but their simulations did not actually contain SMBHs, because the leap-frog time integration scheme (standard for collisionless codes) and the need for gravitational softening prevented the possibility of accurately tracking the close encounters with SMBHs. 
\citet{Karl2015} and \citet{Rantala2017} introduced hybrid codes that combine the tree-code approach with regularization techniques suitable for accurate computation of SMBH orbits and stellar encounters with SMBHs. Although the force errors in the conventional tree-code schemes are too large for reliable analysis of loss-cone orbits, as demonstrated in the appendix of \citet{Gualandris2017}, these approaches seem promising. A combination of the fast-multipole method and regularization techniques is conceivable, and would be the best method for studying the evolution of such stellar systems.

A very different approach to studying the dynamical interactions between stars and a binary SMBH is offered by scattering experiments, which determine the statistical properties of a large ensemble of independent three-body interactions \citep[e.g.][]{Quinlan1996,Sesana2006,Sesana2008,Chen2008}. This information, augmented by suitable assumptions about the distribution of incoming stars, could be used to approximately follow the evolution of the binary and estimate the TDE rates \citep{Sesana2007,Sesana2010,Chen2011,Darbha2018}. The main limitation of this approach is the lack of self-consistency regarding the distribution of stars: it either remains fixed or is evolved using simplified prescriptions, in contrast to a self-consistent evolution in $N$-body simulations (and, of course, in real galaxies).

In this work we use a hybrid Monte Carlo approach \citep{Vasiliev2015a,Vasiliev2015b} that combines the benefits of scattering experiments and $N$-body simulations, while largely avoiding their respective limitations.
The evolving stellar system is still represented by $N$ particles that do not correspond to individual stars, but rather sample their distribution function probabilistically. These particles move in the smooth gravitational potential, self-consistently computed from the ensemble of particles, and can interact with the binary SMBH.

Unlike conventional $N$-body simulations, the gravitational potential is expressed in terms of a spherical-harmonic expansion, centered on the binary centre-of-mass, with the radial dependence of each multipole term in this expansion described by a smooth (spline) curve with a relatively small ($\sim20-30$) number of grid points in $\log r$. This resembles the self-consistent field method of \citet{HernquistOstriker1992}, with the difference that we use spline interpolation instead of a weighted combination of basis functions in radial coordinate. The `MEX' method of \citet{Meiron2014} also computes the radial dependence of each multipole term directly from particle coordinates, without decomposing it into the basis-set expansion, but only at each particle's position, without constructing a global smooth approximation for the potential.
In our implementation, we first create a smooth approximation to the density profile from particle positions, which is also expressed in terms of a spherical-harmonic expansion with radially-interpolated coefficients, and then solve the Poisson equation by numerically evaluating the 1d integrals in radius for each multipole term (for details see \citealt{Vasiliev2018}, Section A.3.1).

The potential of the entire system is updated not at every timestep, as in ordinary $N$-body simulations, but much less frequently -- at the end of each `episode', which should be much shorter than the characteristic timescale of the global evolution (e.g., the time needed to significantly shrink the binary), but still could be longer than the dynamical time, at least for the most bound particles in the central part of the system. Accordingly, the particles move in a smooth static stellar potential for the entire episode, but they do experience the time-dependent forces from the two SMBHs, which themselves follow a Keplerian orbit with fixed parameters during each episode. The time-dependent (non-monopole) part of the binary potential is only important when a particle is close enough to the centre. We record the changes in energy and angular momentum of all particles entering and exiting the sphere of radius $r_\mathrm{enc} = 5a$ (where $a$ is the semimajor axis of the binary). At the end of each episode, these changes are summed up for all particles that had close encounters with the binary, and the binary orbit ($a$ and eccentricity $e$) is adjusted by subtracting the equivalent amount of energy and angular momentum, and optionally taking into account the GW losses.
Hence our approach may be viewed as a superposition of individual three-body scattering experiments, but with the parameters of the incoming stars drawn from their actual distribution function, which evolves in the course of the simulation.

Because each particle's orbit is computed independently from the others, the computational cost scales as $\mathcal{O}(N)$, and the method is trivially parallelized (we use the \textsc{OpenMP} approach to distribute the load to all cores of a single workstation).
Despite this favourable scaling, the number of particles in our runs is typically $\lesssim 10^6$, similar to direct $N$-body simulations. However, unlike the latter, our method has a dramatically lower ($\sim\!100\times$, see Figure~1 in \citealt{Vasiliev2015a}) level of two-body relaxation for the same number of particles, due to several factors.
First, the smooth global potential, with a low number of spherical-harmonic terms and a rather coarse radial grid, acts as a low-pass filter (similar to the use of a relatively large but spatially variable softening length, avoiding a loss of force resolution at small radii). Second, in updating the potential, we do not just use the instantaneous positions of particles at the end of each episode, but rather take $N_\mathrm{samp}\gg 1$ points sampled from each particle's orbit during the entire episode, further suppressing the discreteness noise. Finally, the rather long update intervals for the potential act as a temporal smoother.

Since the loss-cone problems involve both collisional and collisionless processes, we also need to simulate the effect of two-body relaxation. This is achieved by adding perturbations to particle velocities at every timestep, with the magnitude determined by diffusion coefficients computed from the distribution function of all stars, which is also updated after each episode. Importantly, the amplitude of these perturbations can be assigned manually to mimic the level of relaxation expected in a stellar system with $N_\star$ stars, which is unrelated to the actual number of particles $N$ in the simulation. Various tests performed in previous studies demonstrate a good agreement between the Monte Carlo and $N$-body simulations of the same system, when the level of relaxation in the former approach is set to match the latter.

\citet{QuinlanHernquist1997} and \citet{Hemsendorf2002} developed hybrid codes combining the representation of the stellar potential in terms of a multipole expansion with a direct $N$-body simulation of a small number of particles in the very centre, including the two SMBHs. Our approach resembles these methods (especially in the first part), but with important differences (additional temporal smoothing and oversampling to reduce the level of noise, and the explicit account for two-body relaxation).

The Monte Carlo code \textsc{Raga}, introduced in \citet{Vasiliev2015a}, was used to study the TDE rates by single SMBHs in \citet{Vasiliev2014a, LezhninVasiliev2016}, and applied to the binary SMBH evolution in \citet{Vasiliev2015b}. The new version of the code, used in the present paper, is built on the same principles, but redesigned almost from scratch; it is more robust and computationally efficient, and can simultaneously deal with both the binary SMBH evolution and the captures of stars by each SMBH. The code is publicly available as part of the \textsc{Agama} library for galactic dynamics \citep{Vasiliev2019}.

The main advantages of the Monte Carlo approach in the context of dynamics of galactic nuclei are:
\begin{itemize}
\item Correct balance between collisional and collisionless effects (unlike $N$-body simulations with a scaled-down number of particles);
\item Self-consistent treatment of the evolution of the stellar distribution and the loss-cone repopulation (unlike scattering experiments).
\item Possibility of using physically correct size of the loss cone (unlike \citet{Li2017}, who had to rely on extrapolation).
\item Moderate computational cost -- simulations with $10^6$ particles run for a significant fraction of the Hubble time only take about a day on a conventional 16-core workstation.
\end{itemize}
The limitations of the method:
\begin{itemize}
\item It can only be applied to systems with a well-defined centre and an already formed hard SMBH binary, so we cannot reliably simulate the early stages of evolution, when the TDE rates briefly reach a peak \citep[e.g.,][]{Chen2009,Li2017}.
\item We only follow the evolution of binary semimajor axis and eccentricity, and neglect the orientation; however, the latter is not expected to change dramatically, unless the stellar cluster is strongly rotating and the binary orbital plane is misaligned with it \citep{RasskazovMerritt2017}.
\item We also assume that the binary resides in the galaxy centre, neglecting the Brownian motion. \citet{Bortolas2016} found that allowing the SMBH to wander has little effect on the evolution of systems where the loss cone is replenished mainly by collisionless processes. On the other hand, \citet{HolleyKhan2015,Mirza2017} discovered a regime where the SMBH centre-of-mass exhibits coherent motion around the galactic nucleus, which occurs when the stars in the nucleus are on orbits corotating with the binary; however, this does not seem to significantly affect the hardening rate.
\item Our approach for simulating two-body relaxation does not account for its possible enhancement due to resonant relaxation \citep{RauchTremaine1996}; however, we do not expect it to play any role in highly chaotic three-body interactions.
\item Stellar binaries are absent in our simulations; in reality they may be an important source of hypervelocity stars, as predicted by \cite{Hills1988} for single SMBH--binary star systems and by \cite{Wang2018} for binary SMBH--binary star systems.
\item We do not explore the effect of mass segregation (all particles have the same dynamical mass); however, any pre-existing segregation is expected to be erased during the merger \citep{GualandrisMerritt2012}, and is unlikely to be re-generated within the lifetime of the binary (although this is difficult to avoid in $N$-body simulations, see e.g.\citealt{Khan2018}).
\end{itemize}

\section{Variants of models}  \label{sec:ic}

We explore a large suite of models, varying the parameters of the binary SMBH and the stellar system. We only consider `isolated' galaxies, initially constructed to be in equilibrium, not the merger products: \citet{Vasiliev2015b} and \citet{Gualandris2017} demonstrated that the long-term evolution of the binary depends qualitatively on the geometry of the stellar potential (spherical, axisymmetric or triaxial), but only moderately on its particular properties such as axis ratios \citep{Bortolas2018b}, and is similar between isolated and merger simulations.
Therefore, we examine three series of models, with initial profiles following the \citet{Dehnen1993} model having the inner slope $\gamma=1$ and constant axis ratios -- $1:1:1$ for spherical, $1:1:0.8$ for axisymmetric and $1:0.9:0.8$ for triaxial models. These models are designed to be in a self-consistent equilibrium with a single SMBH in the centre, which has a mass 0.01 times the total stellar mass. The initial $N$-body snapshots are prepared with the Schwarzschild orbit-superposition code \textsc{Smile} \citep{Vasiliev2013} and contain $0.5\times10^6$ particles. Next we take 25\% of particles with lowest values of angular momentum, and replace each one with 5 identical particles of a correspondingly smaller mass, bringing the total number of particles to $N=10^6$. This static mass refinement scheme (similar to the one in \citealt{LezhninVasiliev2016}) improves the statistics of TDE rates, since the more numerous smaller particles better sample the underlying distribution function (due to independent velocity perturbations, they quickly spread out in the phase space). Thus our mass resolution is equivalent to a $2.5\times10^6$ particle snapshot in the low-angular-momentum part of the phase space, which is responsible for the interaction with the SMBH binary and for the TDE rates. The presence of different mass groups does not lead to mass segregation, since we use the same the drift coefficient in the treatment of two-body relaxation for all particles (i.e., particle have different tracer masses but identical dynamical masses).

The single SMBH is then replaced by a binary with the same total mass $M_\bullet$ and a semimajor axis $a$ equal to that of a `hard' binary \citep[Equation 8.71]{MerrittBook}:
\begin{align}  \label{eq:ah}
a_\mathrm{h} \equiv r_\mathrm{infl} \frac{q}{4\,(1+q)^2},
\end{align}
where $r_\mathrm{infl}$ is the influence radius initially containing the stellar mass equal to $2M_\bullet$ (for our initial conditions, $r_\mathrm{infl}=0.165\,N$-body length units), and $q\le 1$ is the mass ratio of the two SMBHs.
We explore a range of initial binary eccentricities ($e=0.01, 0.3, 0.6$ and $0.9$) and mass ratios ($q=1, 1/3, 1/9$, $1/27$ and $1/81$). The models are scaled to physical units in the following way: we consider three values for the total mass of the binary $M_\bullet = 10^7, 10^8$ and $10^9\,M_\odot$, and assign the length scale from the requirement that 
\begin{align}  \label{eq:rinfl}
r_\mathrm{infl} = 35\,\mbox{pc}\times \big[M_\bullet / 10^8\,M_\odot\big]^{0.56}
\end{align}
\citep[Equation 2.16]{MerrittBook}. We caution that this relation effectively reduces the diversity of galactic nuclei to a one-parameter family determined by $M_\bullet$, whereas in reality there is a considerable scatter of $M_\bullet$ values for galaxies of a similar mass, concentration, or central density slope. Hence our results should be viewed as average trends rather than specific predictions, bearing in mind that merger timescales and capture rates could vary by a factor of few for a given $M_\bullet$. As the binary SMBH erodes the central density cusp at the early stage of evolution, the influence radius increases by up to 20\% (for $q=1$).

The loss-cone radius for each SMBH is set by Equation~\ref{eq:rcapt}; particles reaching a smaller distance to the SMBH are captured and removed from the simulation, with their mass added to the respective SMBH mass (although the total added mass is always significantly smaller than the initial SMBH mass). The relaxation level is set by the number of stars in the physical system being modelled, using the value of Coulomb logarithm $\ln\Lambda = \ln(0.3 M_\bullet / M_\odot)$. We stress that the number of particles in the simulation is always the same ($10^6$), but we quote all physical values (e.g., the galaxy mass) according to the chosen dimensional scaling factors.
The simulations are run for several thousand $N$-body time units, corresponding to a physical time of up to 3 Gyr. 

To compare the TDE rates between galactic nuclei hosting single and binary SMBHs, we also conducted simulations with single SMBHs. The complication in such a comparison is that a binary SMBH disrupts the pre-existing stellar cusp at the early stage of its evolution, resulting in a shallower density profile. To account for this, we started the single SMBH runs from the snapshots of corresponding simulations with the binary SMBH after the initial stage of relatively fast evolution ($\sim10^7$~yr). After this early stage, the density profile changes very little in the simulations with binary SMBHs, so the comparison is more adequate.

Given the large number of parameters (total mass of the binary $M_\bullet$, mass ratio $q$, initial eccentricity $e$, three variants of geometry, inclusion or absence of captures), we do not discuss all possible combinations of them, but instead focus on the main trends arising from changing one parameter at a time. In general, we believe the triaxial geometry to be the most relevant physically, given that a realistic galaxy merger would not produce a precisely axisymmetric (much less spherical) system, but we include the other two cases for comparison.

\section{Results}  \label{sec:results}

\subsection{Hardening rate of the binary}  \label{sec:hardening}

\begin{figure}
    \includegraphics{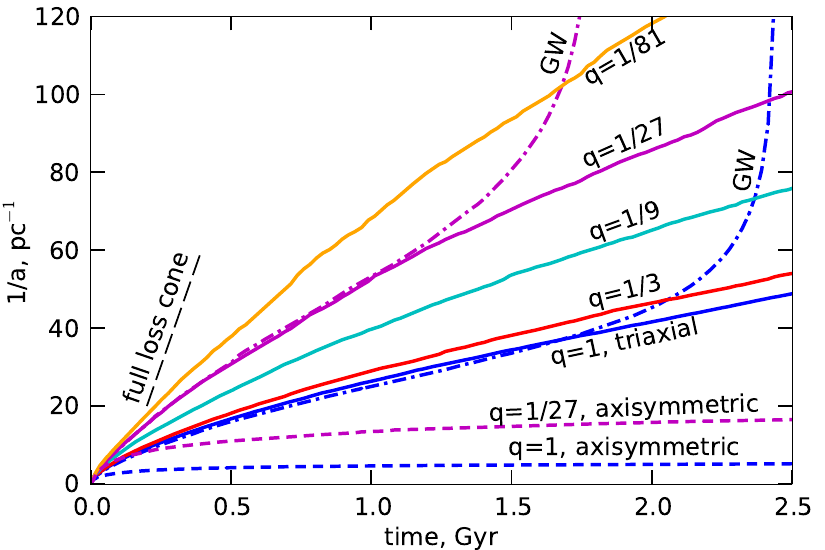}
    \caption{Evolution of inverse semimajor axis of the binary $1/a$ with total mass $M_\bullet=10^8 M_\odot$ and mass ratio $q=1$ (blue), $q=1/3$ (red), $q=1/9$ (cyan), $q=1/27$ (magenta) and $q=1/81$ (orange). 
    Solid lines show the triaxial systems, which happily continue hardening, and dashed -- axisymmetric systems, which virtually stall after a short period of initial hardening, due to the depletion of the loss cone (spherical models, not shown on this plot, stall even sooner).
    Dot-dashed lines show equivalent simulations with GW emission turned on, which reach coalescence in $\sim2-2.5$~Gyr in the case of zero eccentricity; for a higher initial eccentricity this would occur sooner. Dashed black line illustrates the hardening rate expected in the initial model if the loss cone were full (Equation~\ref{eq:fulllc}), which is clearly much higher than the actual rate measured in simulations. This reference full-loss-cone rate is somewhat reduced at later stages due to a partial destruction of the cusp, but the actual hardening rate still remains considerably lower because of the depletion of centrophilic orbits, which is less prominent for $q\ll 1$.
    }
\label{fig:sma}
\end{figure}

The hardening rate is $S = \mathrm{d}(1/a) / \mathrm{d}t$ is usually compared to the full-loss-cone hardening rate:
\begin{subequations}  \label{eq:fulllc}
\begin{align}
S_\mathrm{full} \equiv H\,G\,\rho / \sigma,
\end{align}
where $\rho$ and $\sigma$ are the density and velocity dispersion of stars in the galactic nucleus, and $H\sim 15$ is a dimensionless coefficient calibrated by scattering experiments. The practical difficulty with this definition is that the galactic nucleus is not a uniform-density background, but \citet{SesanaKhan2015} suggested that the values of $\rho$ and $\sigma$ taken at the influence radius (\ref{eq:rinfl}) provide a reasonable estimate.
Another possible definition (\citealt{MerrittBook}, Equations~8.111--8.115, or \citealt{Vasiliev2015b}, Equation~4) is
\begin{align}
S_\mathrm{full} \equiv (2\pi)^{3/2}\,H\,G \int f(E)\,\mathrm{d}E,
\end{align}
\end{subequations}
where the integration is carried over the range of stellar energies where the stars are expected to efficiently repopulate the loss cone. One may take the value of the stellar potential at origin $\Phi_\star(r=0)$ as the lower limit of integration, which is equivalent to considering only the stars that are unbound to the SMBH (but of course still bound to the entire galaxy). These two definitions typically agree to within a factor $\lesssim 1.5$.

Of course, the stellar density profile itself evolves with time (see Section~\ref{sec:massdef}). The early stage of binary formation and hardening leads to a rapid erosion of the pre-existing stellar cusp \citep[e.g.,][]{MilosMerritt2001}, and afterwards the density continues to decrease, but very gradually. Most of the stars that interact with the binary at the later stage are scattered into the loss cone from relatively large radii, hence their disappearance has little effect on the density profile. Accordingly, we compute the value of the `full-loss-cone' hardening rate right after the formation of the hard binary, and neglect its subsequent evolution.

In agreement with other studies, we find that the hardening rates have little dependence on eccentricity (less than 10\%). Of course, the actual lifetime of the binary strongly depends on $e$, because the efficiency of GW emission is proportional to $(1-e^2)^{-7/2}$ \citep[e.g.,][Equation~4.234]{MerrittBook}. In most of our simulations, we turn off the GW losses and focus on purely stellar-dynamical effects in the binary evolution, bearing in mind that once the semimajor axis reaches a critical value 
\citep[Equation~2]{Dotti2012}
\begin{align}  \label{eq:agw}
a_\mathrm{GW} \equiv 0.064\mbox{ pc} \times
\left(\frac{M_\bullet}{10^8\,M_\odot}\right)^{3/4}\!\! \frac{q^{1/4}}{(1+q)^{1/2}\,(1-e^2)^{7/8}} ,
\end{align}
the binary will coalesce in less than $10^{10}$~yr due to the GW emission alone. Since the stellar-dynamical hardening continues to operate, the remaining lifetime of the binary after this moment is comparable to the time needed to reach $a_\mathrm{GW}$ -- typically less than a gigayear.

Figure~\ref{fig:sma} shows the evolution of $1/a$ for models with $M_\bullet = 10^8 M_\odot$ (similar results were obtained for other $M_\bullet$ values) and different $q$. The hardening rate is nearly independent of the eccentricity (the figure shows the runs with $e=0$) and is insensitive to whether the stars are captured by individual SMBHs or not. The likely explanation is that the geometric size of the loss-cone of a single SMBH is much smaller than the cross-section for strong three-body interactions with the binary, therefore the probability of being captured during a single three-body interaction is rather minor, and most stars are ejected and contribute to the binary hardening.

The hardening rate measured in these simulations is substantially lower than the reference value $S_\mathrm{full}$ because of the deficit of particles with low angular momentum. In spherical and axisymmetric systems, the loss cone is repopulated mainly by two-body relaxation, i.e., very slowly. Even in triaxial systems, the population of centrophilic orbits that can enter the loss cone is gradually depleted, hence the hardening rate decreases with time. Nevertheless, the semimajor axis reaches the critical value $a_\mathrm{GW}$ (\ref{eq:agw}) in less than 1~Gyr and continues to shrink; additional simulations with GW emission turned on (shown in dot-dashed line) demonstrate that the actual lifetime of the binary after this moment is much shorter than the Hubble time.
The time-dependent hardening rates in the present study are well described by Equation~21 and Table~1 in \citet{Vasiliev2015b}. A simple and reasonably realistic estimate is to approximate the hardening rate as a constant and take $S\simeq \mu S_\mathrm{full}$, with the dimensionless coefficient $\mu \sim 0.1 - 0.3$ (smaller $q$ values correspond to higher $\mu$). Using the expressions from Section~4.4 in \citet{Vasiliev2015b} and the time-dependent hardening rates measured in our simulations, we compute the expected coalescence time for a range of initial binary eccentricities $e_0$; the results are reasonably well fitted by a simple expression
\begin{align}  \label{eq:tcoal}
T_\mathrm{coal} \approx 1\mbox{ Gyr} \times
\left(\frac{M_\bullet}{10^8 M_\odot} \right)^{-1}
\left(\frac{r_\mathrm{infl}}{30\mbox{ pc}} \right)^{2}
\left(\frac{\mu}{0.1} \right)^{-4/5}
(1-e_0) ,
\end{align}
with a weak dependence on $q$ which may be ignored. Note that we have used a scaling relation (\ref{eq:rinfl}) linking $M_\bullet$ to $r_\mathrm{infl}$ in a specific way; the two quantities appear independently in the above formula following the analytical considerations of \citet{Vasiliev2015b}. 
For our choice of slope and normalization of the $M_\bullet$ -- $r_\mathrm{infl}$ relation, the timescale is nearly independent of $M_\bullet$ and lies in the range $2-2.5$~Gyr for $e_0=0$. We recall that the influence radius is somewhat increased from its initial value during the long-term evolution, hence our timescales are $\sim2\times$ longer than those in \citealt{Vasiliev2015b}. Moreover, models with $q\ll 1$ suffer less from the cusp destruction at early stages, and hence their influence radius during most of the evolution is somewhat smaller than in models with $q\simeq 1$; the statement about weak dependence of $T_\mathrm{coal}$ on $q$ takes into account this difference in $r_\mathrm{infl}$.

For spherical and axisymmetric systems, the evolution of $1/a$ slows down dramatically after the initial phase, and $a$ barely reaches $0.1\,a_\mathrm{h}$, which is still far larger than $a_\mathrm{GW}$. Therefore, unless $e_0$ is close to unity or other non-stellar-dynamical processes are considered, the binary SMBHs would not merge in these cases, in agreement with \citet{Vasiliev2015b,Gualandris2017}. Using direct $N$-body simulations, \citet{Khan2013} argued that the hardening rate remains sufficiently high even in axisymmetric systems to reach coalescence within the Hubble time, which seems to contradict our findings. However, their different conclusion is likely to result from the enhanced two-body relaxation in the simulations with $N\sim 10^6$ compared to real galaxies with $N_\star \sim 10^8-10^{10}$, as discussed earlier and corroborated by the analysis of $N$-dependence of the hardening rate in the two papers quoted above.

\subsection{Eccentricity evolution}  \label{sec:ecc}

\begin{figure}
    \includegraphics{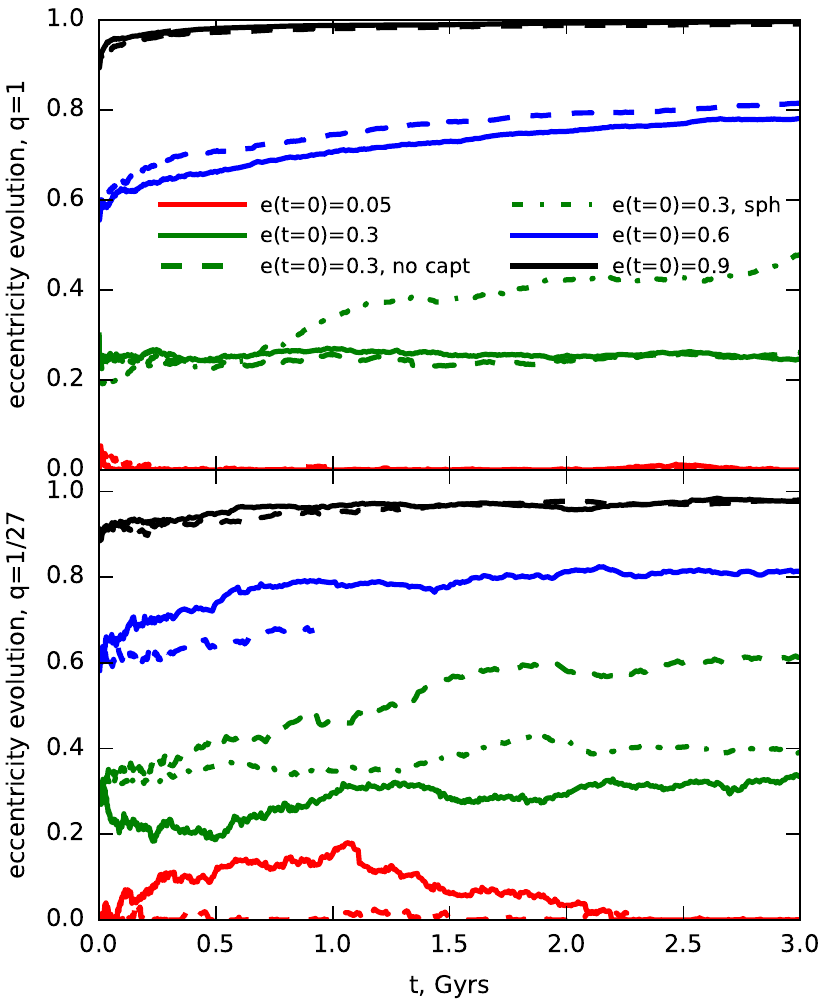}
\caption{Eccentricity evolution over time for simulations with initial eccentricities equal to $e=0.01$ (red), 0.3 (green), 0.6 (blue), 0.9 (black) in a triaxial potential with $M_{\bullet}=10^9 M_\odot$ and mass ratios $q=1$ (upper panel) and $q=1/27$ (lower panel). We also demonstrate how the eccentricity evolves in the spherical potential (dot-dashed line). Dashed lines correspond to simulations without captures. In all cases with $e>0.3$ the eccentricity growth and saturation is evident. However, the overall evolution of $e(t)$ is quite stochastic, especially for $q\ll 1$.}
\label{fig:ecc}
\end{figure}

\begin{figure}
    \includegraphics{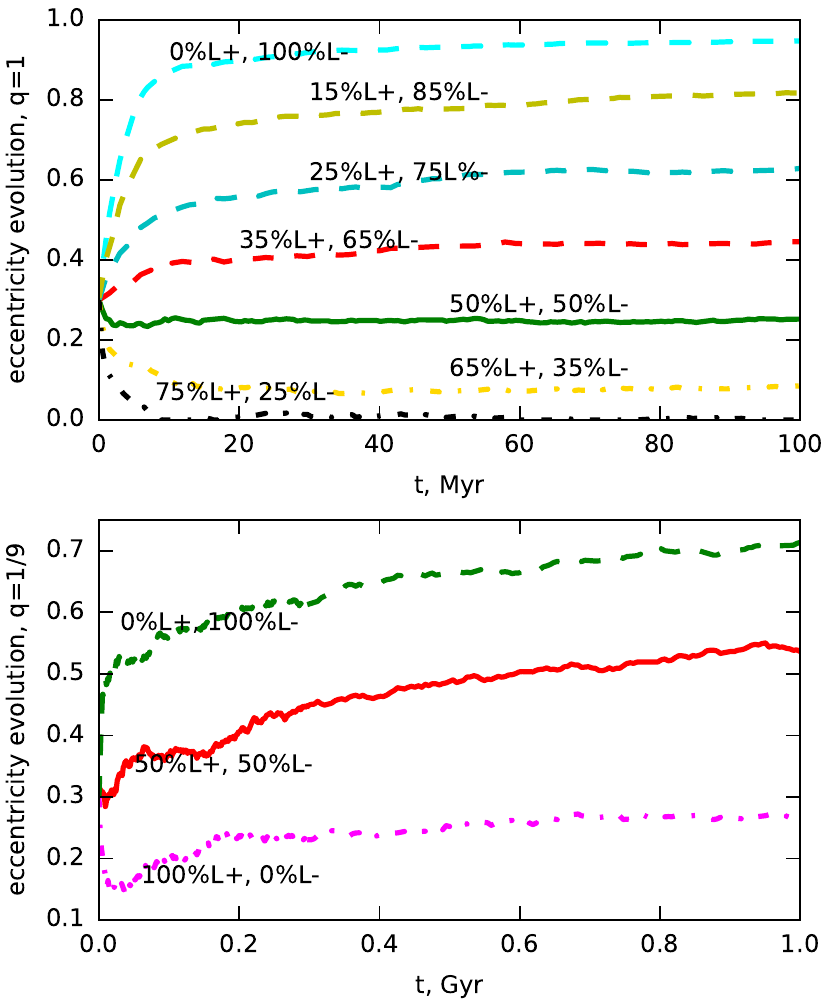}
    \caption{Eccentricity evolution over time for simulations with the initial eccentricity equal to $e=0.3$, mass ratios $q=1$ (upper panel) and $q=1/9$ (lower panel), and various amount of rotation of the stellar cluster. Solid line represents an isotropic cluster in $L_z$, dashed -- counterrotating clusters with specific fractions of the co- and counterrotating orbits, dot-dashed -- corotating clusters. After the short initial phase of rapid eccentricity growth (in counterrotating clusters) or decay (in corotating ones), the subsequent evolution is much more gradual in both $q=1$ and $q=1/9$ cases, although the long-term growth of eccentricity is more prominent for $q=1/9$.}
\label{fig:eccrotation}
\end{figure}

The eccentricity of the binary changes rather moderately in the case of a (nearly-)isotropic distribution of field stars (Figure~\ref{fig:ecc}).
For models with a rather high initial value ($e\gtrsim 0.6$), it tends to increase further, reaching $e \gtrsim 0.98$ for an initial value $e=0.9$, which, of course, substantially shortens the lifetime of the binary before the GW-induced coalescence. On the other hand, for lower initial values the eccentricity remains roughly constant. The inclusion or neglect of captures has little influence on the eccentricity evolution. However, the geometry of the stellar potential seems to play some role, in particular, the eccentricity grows faster in spherical than in the triaxial cases, though it is clear from Figure~\ref{fig:sma} that the degrees of binary hardness in different geometries are different.

Observationally, there is growing evidence that nuclear star clusters have a substantial amount of rotation (e.g., \citealt{Seth2008}, \citealt{Schoedel2009}, \citealt{Hartmann2011}). This is important for the eccentricity evolution, since the scattered stars coming from prograde (retrograde) orbits tend to increase (decrease) the angular momentum of the binary.
\citet{Iwasawa2011} and \citet{Sesana2011} studied the evolution of very unequal mass ratio binaries ($q \sim 10^{-2}$) and found that the eccentricity strongly grows or declines when the stellar cluster is counterrotating or corotating with the binary orbit. \citet{HolleyKhan2015} and \citet{Mirza2017} confirmed these trends with higher-resolution $N$-body simulations for a wider range of parameters, and additionally found a regime where the binary centre-of-mass settles into a nearly circular orbit with a radius $\sim r_\mathrm{infl}$ in a corotating stellar cluster. The latter study also considered the evolution of binary orbital plane orientation in their simulations, while \citet{RasskazovMerritt2017} addressed this question using Fokker--Planck formalism.

To explore the impact of the net rotation of the stellar cluster, we conducted an auxiliary set of simulations with the initial distribution of stars modified in the following way. To bias the stellar distribution towards corotation (counterrotation), we change the sign of the velocity for particles with negative (positive) $z$-component of angular momentum, with a probability ranging from 0 to 100\%. This preserves the orbital structure of the stellar system (required for it to remain in a self-consistent equilibrium) and a nearly-\-isotropic distribution of orbit eccentricities (an assumption used in computing the two-body relaxation rate in the Monte Carlo approach).

Figure~\ref{fig:eccrotation} shows the evolution of binary eccentricity in a range of triaxial $M_\bullet=10^9$, $q=1$ or $1/9$ systems with different degree of rotation, starting from the initial value $e=0.3$. In agreement with previous studies, the eccentricity rapidly decreases or increases in corotating or counterrotating clusters, correspondingly. However, this stage lasts only around $10^7$~yr, after which the eccentricity reaches a nearly stationary value for $q=1$, or keeps increasing only moderately for $q=1/9$. The likely reason is that the loss cone is repopulated mainly from centrophilic orbits on longer timescales, and they do not conserve the sign of angular momentum, hence the orbits of incoming stars are eventually isotropized. 

In the absence of GW losses, the hardening rates are not significantly affected by eccentricity. The capture rate is moderately (less than a factor of two) suppressed for corotating systems. The inclusion of captures slightly enhances the increase of eccentricity in the majority of our simulations, as predicted by \citet{Sesana2008, Chen2011}. In the case of very unequal-mass mergers, the eccentricity at the time of the formation of a hard binary is expected to be higher, having grown at the previous stage of dynamical friction \citep{DosopoulouAntonini2017}.

\subsection{Capture rates}  \label{sec:capture}

\begin{figure*}
    \includegraphics{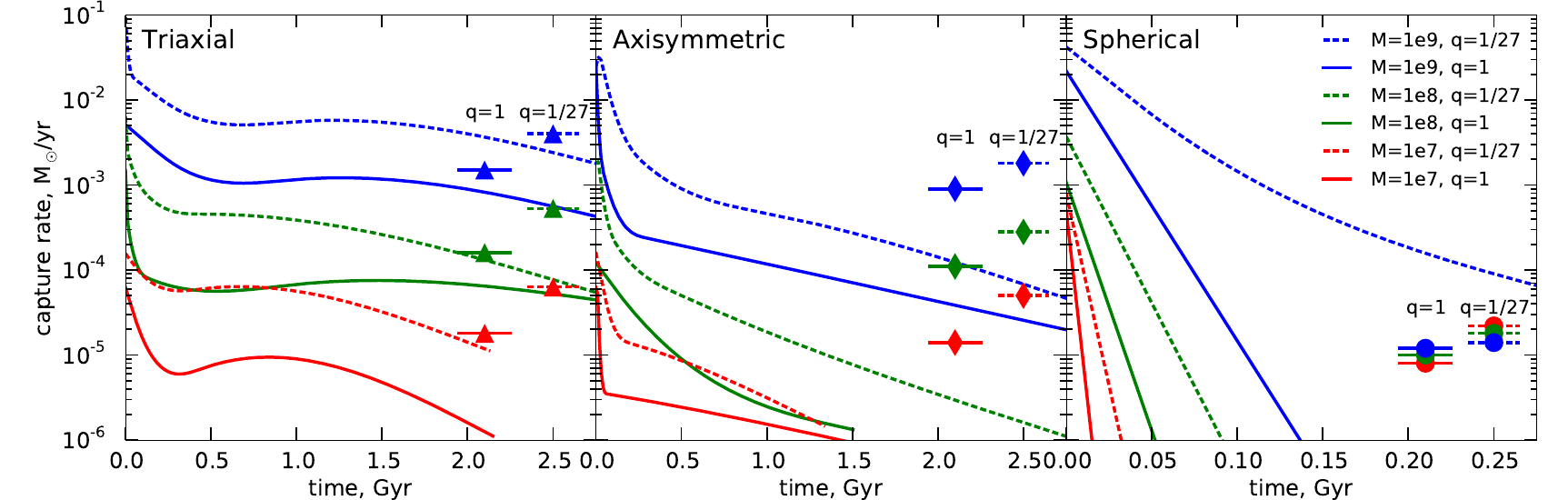}
    \caption{Time-dependent capture rates in simulations of binary SMBH in various geometries: triaxial (left panel), axisymmetric (centre), spherical (right). Colour corresponds to the SMBH mass, from top to bottom: $10^9$ (blue), $10^8$ (green) and $10^7~M_\odot$ (red); solid lines show the equal-mass case ($q=1$), and dashed -- for the mass ratio $q=1/27$. 
    Average capture rates in otherwise equivalent galaxies with a single SMBH are shown by symbols in the right side of each panel (they are approximately constant over time). Since the density cusp is less depleted in unequal-mass cases, the rates are higher for $q\ll 1$ than for $q=1$.
    In simulations with binary SMBH, the rates are initially higher, but decline over time (faster in more symmetric potentials -- in a spherical geometry they are essentially zero after a fraction of Gyr, note the different scale of the horizontal axis in the right panel). }
    \label{fig:capture}
\end{figure*}

We now discuss the dependence of capture rates on the parameters of the binary SMBH, and compare them with the rates expected in galactic nuclei hosting a single SMBH. As explained in Section~\ref{sec:losscone}, black holes more massive than a few${}\times10^7\,M_\odot$ swallow most stars entirely (except very extended giants), without producing an observable TDE flare. We refer to all cases when a star passes at $r<r_\mathrm{LC}$ (Equation~\ref{eq:rcapt}) as captures, regardless of the outcome.

Figure \ref{fig:capture} shows the time-dependent capture rates obtained in our simulations of spherical, axisymmetric, and triaxial clusters, for three different SMBH masses ($10^7, 10^8$ and $10^9~M_\odot$) and two values of mass ratio ($q=1$ and $q=1/27$). For comparison, we also plot the average rates in equivalent simulations with a single SMBH, started from the initial configuration shortly after the formation of a hard binary (when the original density cusp has already been destroyed, but the population of loss-cone orbits has not yet been depleted). These ``reference rates'' are typically around few${}\times 10^{-5}~M_\odot\,\mbox{yr}^{-1}$ in spherical systems \citep[e.g.,][]{WangMerritt2004, StoneMetzger2016}, but higher in non-spherical cases (e.g., Figure~4 in \citealt{Vasiliev2014a} or Figure~5 in \citealt{LezhninVasiliev2016}).

As seen from the right panel (spherical case), the capture rates in simulations with binaries are initially much higher than the steady-state rates in galaxies with a single SMBH, but they rapidly decline with time. We caution that this early transient period is rather sensitive to the particular way of setting up the initial conditions, so one should not compare these values quantitatively, but the difference of more than an order of magnitude is a robust feature. This enhancement has been found previously in several studies \citep{Chen2011, WeggBode2011, Li2015}, but lasts only a short time ($\lesssim 10^7$~yr).

The initial transient period is similar in all three geometries, but the late-time evolution is rather different. In the spherical case, both the capture rates and the binary hardening rates drop essentially to zero after this initial period, because the loss cone is nearly empty. In the axisymmetric case the situation is similar but less dramatic; nevertheless, the capture rates drop by more than an order of magnitude compared to the single-SMBH case. In the triaxial case, however, the long-term capture rates are only moderately smaller or even comparable to those found in single-SMBH systems, scaling nearly linearly with the SMBH mass. This proportionality is an indication of the major role of collisionless loss-cone refilling mechanism, whose rate is independent of the number of stars (recall that the geometric size of the loss cone is proportional to $M_\bullet$ for direct-capture events). 

The capture rates are insensitive to the binary eccentricity, varying by less than 30\%. 
In the equal-mass binary ($q=1$), each SMBH captures roughly half of the stars, but the proportion of stars captured by the primary (more massive) component of the binary tends to 100\% very quickly as the mass ratio $q$ decreases (even in the case $q=1/3$, this fraction is $\gtrsim90\%$). On the other hand, since a massive enough primary SMBH would swallow stars entirely, the secondary remains the only source of observable TDE \citep{FragioneLeigh2018}.

Recently \citet{Darbha2018} studied the rates of TDE by binary SMBHs, using scattering experiments (drawing from a uniform distribution of incoming stars in angular momentum) under the assumption of a full loss cone. They found that the rates are increased compared to the case of a single SMBH by a factor of few, weakly depending on the mass ratio (as long as it is higher than few ${}\times10^{-2}$). The likely explanation is that the time-dependent potential of the binary perturbs the incoming orbits and forces a larger fraction of them to enter the loss cone of either of the two SMBH, especially in the case of chaotic resonant scattering events with multiple pericentre passages. We  ran a series of isolated scattering experiments, using a similar setup to \citet{Darbha2018}, and confirmed their conclusions regarding the enhancement of capture rates in binary systems.

At a first glance, this seems to contradict our findings that the capture rates by binaries are smaller than or comparable to those by single SMBHs in simulations of triaxial systems, where the loss cone is at least partially full (the most relevant case for comparison). However, the analysis of time-dependent capture rates in our simulations indicates a different behaviour in single- and binary-hosting nuclei. At the early stage of evolution, the rates are considerably higher for systems with binary SMBHs, but subsequently they decrease; their time evolution is similar to that of the hardening rates, and the slowdown is due to the loss-cone depletion in both cases.
By contrast, in systems with a single SMBH they stay at a roughly constant level, or even increase with time, as the phase-space gap at low angular momenta, carved out by a pre-existing binary, gradually fills up \citep[see][]{MerrittWang2005, LezhninVasiliev2015, LezhninVasiliev2016}. In the case of a continuously evolving binary, this gap is not refilled but rather deepens with time, because most centrophilic orbits that still exist in the system are scattered away by the binary, rather than find their way into the loss cone of a single SMBH. The loss cone of the binary is gradually emptied at any given energy, progressing from inside out to less bound orbits (see the discussion in Section~4.3 and Figure~6 in \citealt{Vasiliev2015b}). The fact that we find comparable capture rates in triaxial single- and binary-hosting nuclei appears to be a coincidence arising from the cancellation of opposite trends -- enhancement of captures by binaries compared to single SMBH for a fixed supply rate of incoming stars, and the decrease of this rate with time in galactic nuclei where the binary continues to evolve.

\citet{Darbha2018} also found that the fraction of stars captured by the secondary SMBH scales roughly linearly with the mass ratio $q$, whereas we observe a much steeper drop in the secondary disruptions as $q$ decreases. The difference likely arises from a shallower dependence of capture radius on the mass of a single SMBH (Equation~\ref{eq:rcapt}) for less massive SMBH ($\lesssim 10^6\,M_\odot$) studied in their work.

\subsection{Loss-cone orbits}  \label{sec:orbits}

\begin{figure}
    \includegraphics{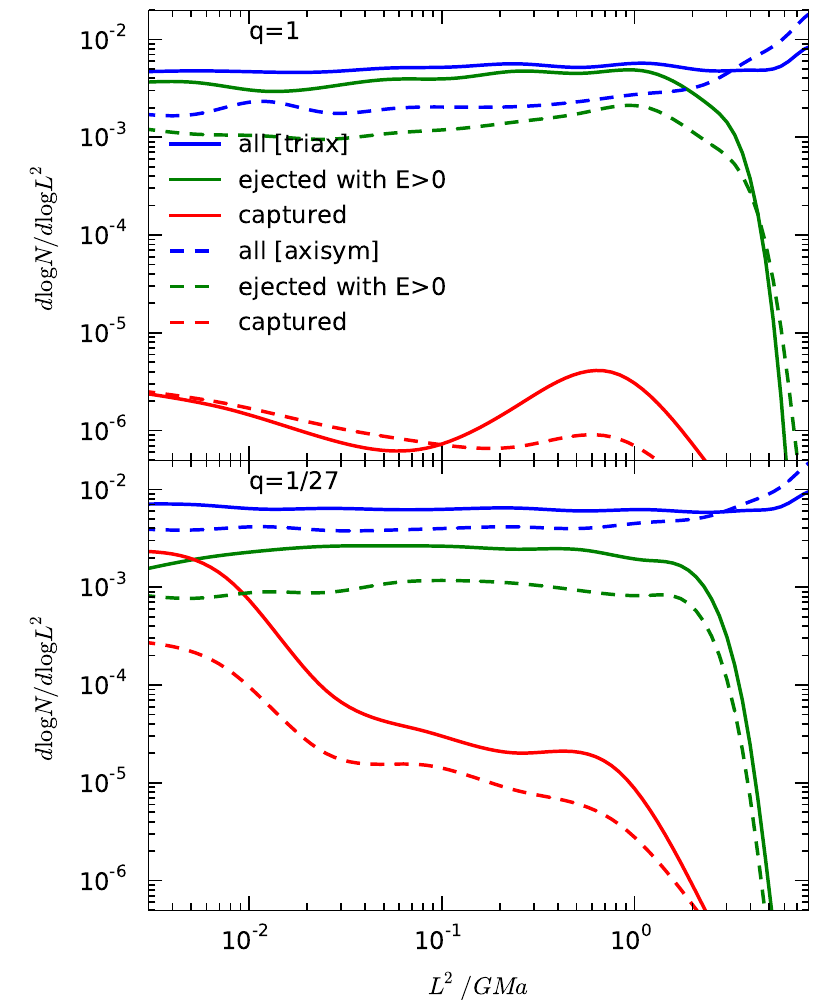}
    \caption{Properties of stars in the loss cone of the binary SMBH. Shown is the distribution of stars in squared angular momentum: if a star were orbiting a single SMBH instead of a binary, this quantity would be a proxy for the pericentre distance.
    Different lines show the outcome of the interaction: blue are all stars that have been scattered by the binary, but remained in the galaxy; green are hypervelocity stars that were ejected from the galaxy with positive total energy; and red are stars that were captured by one of the SMBH. Solid lines are for the triaxial case and dashed -- for the axisymmetric case; top panel is for equal-mass binary, and bottom -- for the mass ratio $q=1/27$. 
    }
    \label{fig:encounters}
\end{figure}

We analyze the properties of stars that interact with the binary in the following way.
For each particle entering or leaving the sphere of radius $5a$ from the binary centre-of-mass, we record the initial and final values of energy and angular momentum, and also the direction of the velocity on exit. Particles that leave the interaction zone with a positive total energy are ejected from the galaxy as hypervelocity stars (see next section); otherwise a particle may return later and experience a `secondary slingshot' \citep{MilosMerritt2003}. Since a three-body interaction may last a long time and consist of multiple close approaches with the binary, we merge the data recorded for successive interactions of the same particle if the time between them is $\le 1$ orbital period at the given energy. The majority of particles experienced more than one interaction, and a substantial fraction of interactions with the initial value of angular momentum $L^2 \lesssim 2GM_\bullet a$ resulted in the ejection of the particle with a positive total energy (Figure~\ref{fig:encounters}). 

In triaxial systems, the distribution of initial values of squared angular momentum (d$N$/d$L^2$) is approximately flat, which is a characteristic sign of a full-loss-cone regime. Note that this does not mean that the interaction rate corresponds to the full-loss-cone rate (\ref{eq:fulllc}): rather, the stars close to the loss-cone boundary are equally likely to cross it with any value of angular momentum, not just the value only slightly smaller than the boundary (as would happen in the empty-loss-cone regime). 
However, in more symmetric systems the situation is different: the dashed lines illustrate that in the axisymmetric case, the probability distribution increases as one moves outwards from the loss-cone boundary (to the right in that plot).
The distribution of captured stars (red lines) is concentrated at low $L^2$ in unequal-mass cases ($q\ll1$), because the more massive component of the binary resides closer to its centre-of-mass. 

The orbits of stars in a time-dependent potential cannot be characterized rigorously (but see \citealt{Li2018} for an attempt); however, it is likely that most of the orbits that bring stars into the vicinity of the binary are centrophilic, i.e., can attain arbitrarily small values of angular momentum (e.g., boxes, pyramids, some minor resonant families, or chaotic orbits fulfill these criteria). At least in the case of a single SMBH in non-spherical system, the majority of captured stars arrive from these orbits \citep[Figure 5 in][]{Vasiliev2014a}, and we expect this to hold also for the loss cone of a binary \citep[see Section~4.2 in][]{Vasiliev2015b}.

\subsection{Hypervelocity stars}  \label{sec:hvs}

\begin{figure}
    \includegraphics{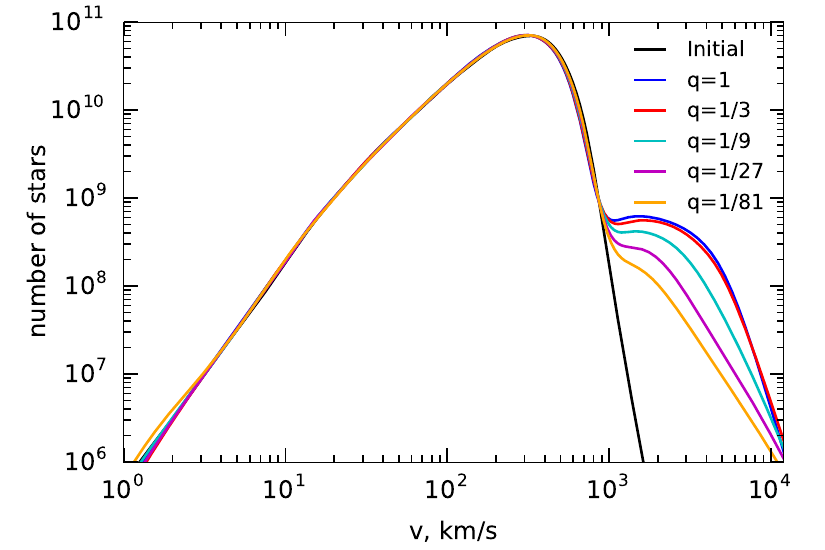}
    \includegraphics{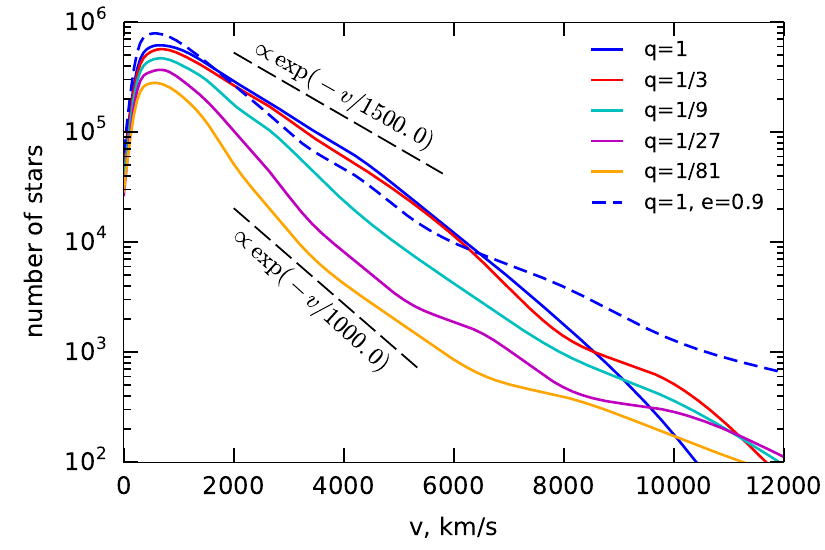}
    \caption{Distribution of stars in velocity after 1~Gyr of evolution, for the case $M_\bullet=10^8\,M_\odot$, $e=0$ and various values of mass ratio: $q=1$ (blue), $q=1/3$ (red), $q=1/9$ (cyan), $q=1/27$ (magenta), $q=1/81$ (orange). In the top panel, all stars are shown (both bound and unbound), and the additional black line corresponds to the initial model. In the bottom panel, only the unbound stars are shown, which roughly follow an exponential distribution; additionally, the case of equal-mass binary on an initially eccentric orbit ($e=0.9$) is shown in dashes, illustrating a longer tail at high ejection velocities. }
    \label{fig:veldist}
\end{figure}

\begin{figure}
    \includegraphics[width=0.49\linewidth]{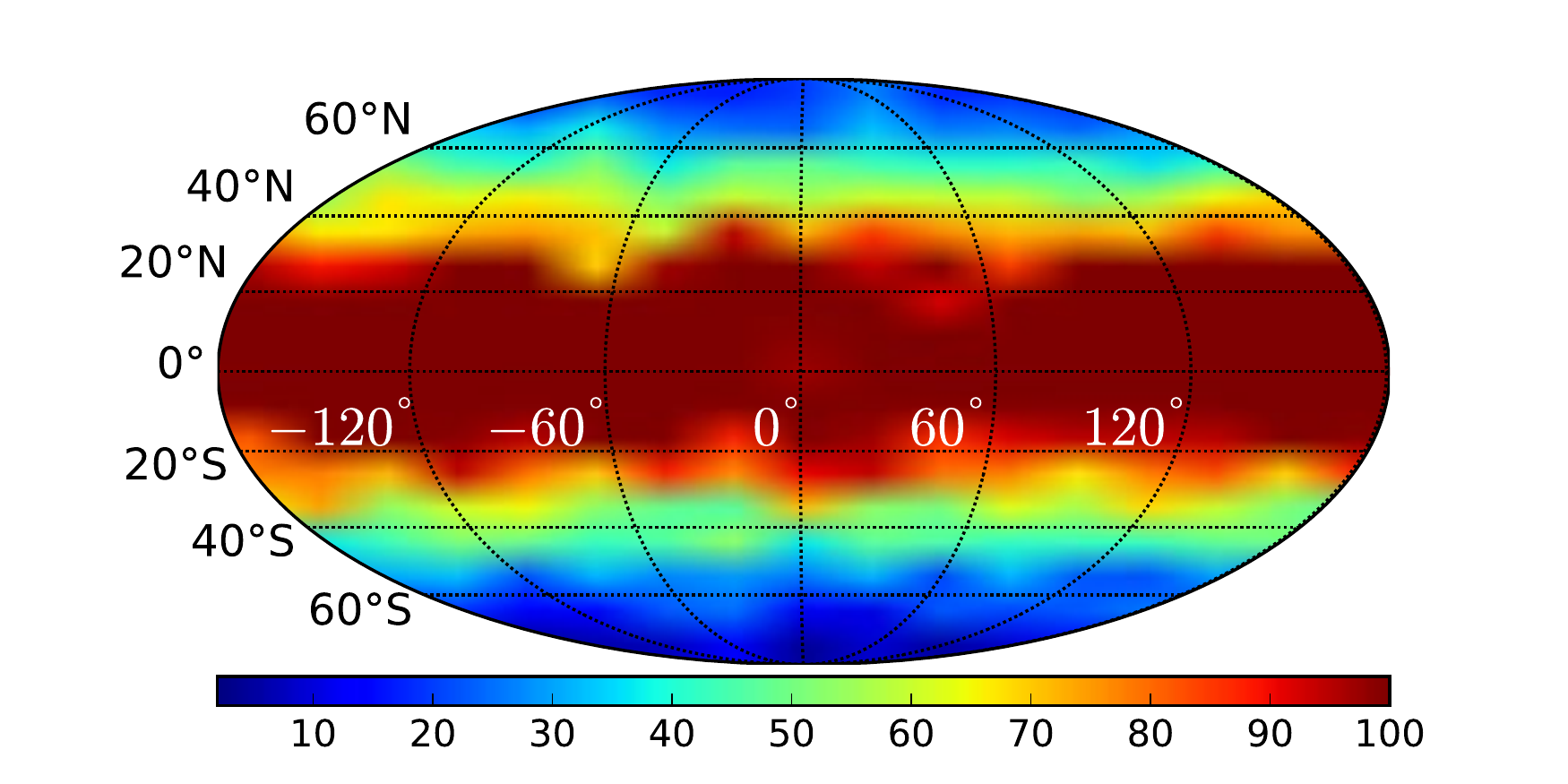}
    \includegraphics[width=0.49\linewidth]{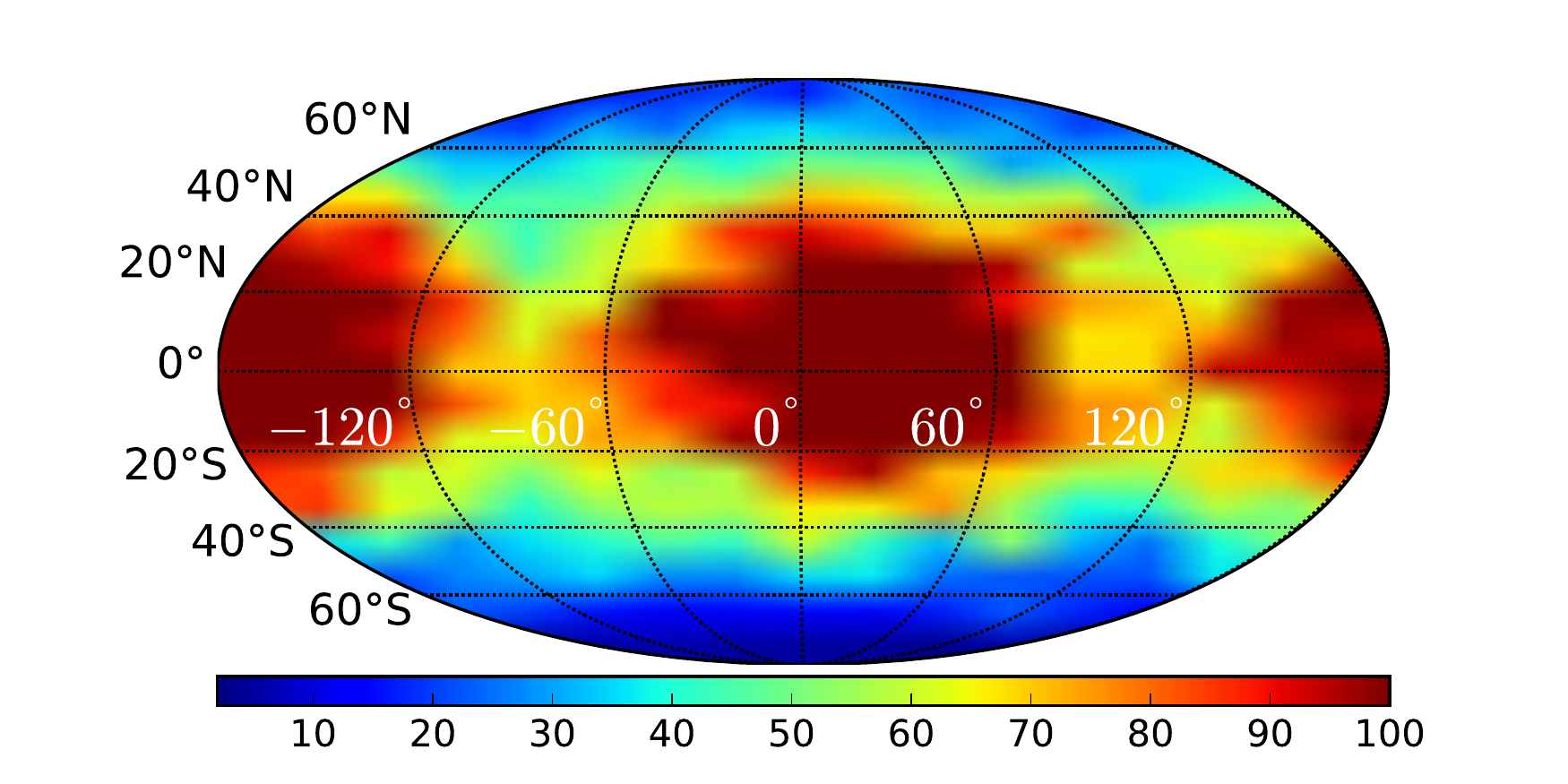}
    \includegraphics[width=0.49\linewidth]{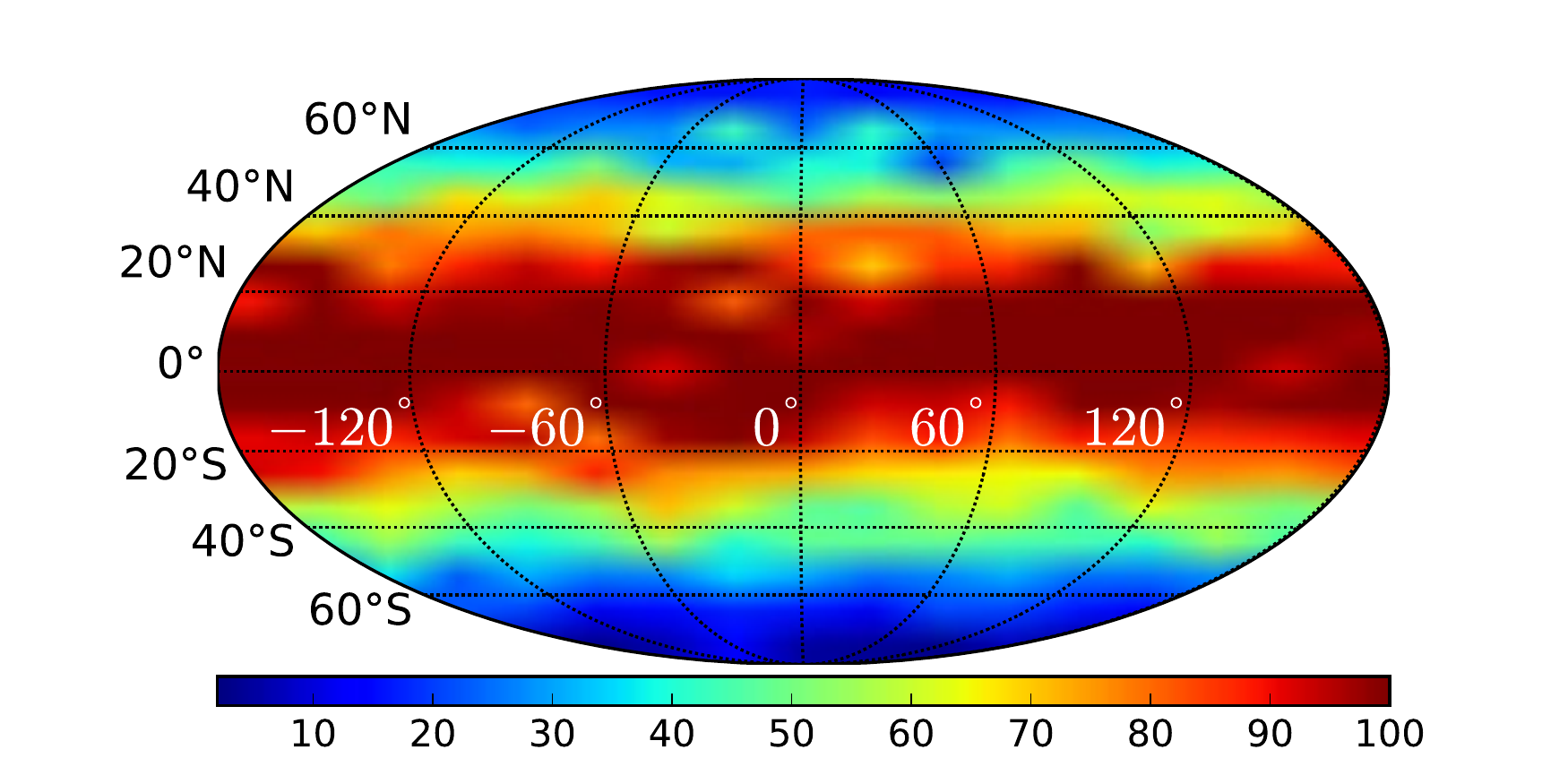}
    \includegraphics[width=0.49\linewidth]{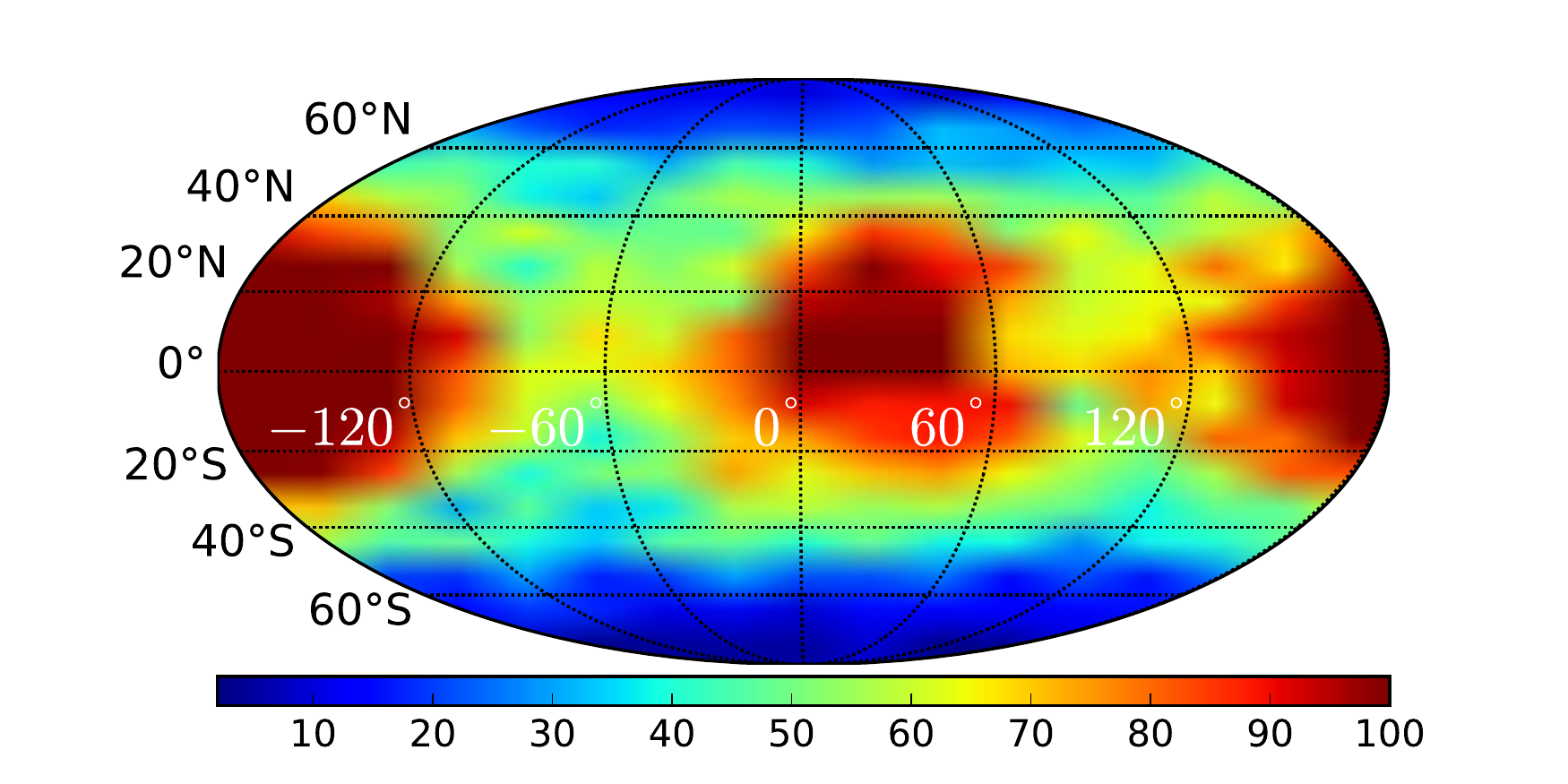}
    \includegraphics[width=0.49\linewidth]{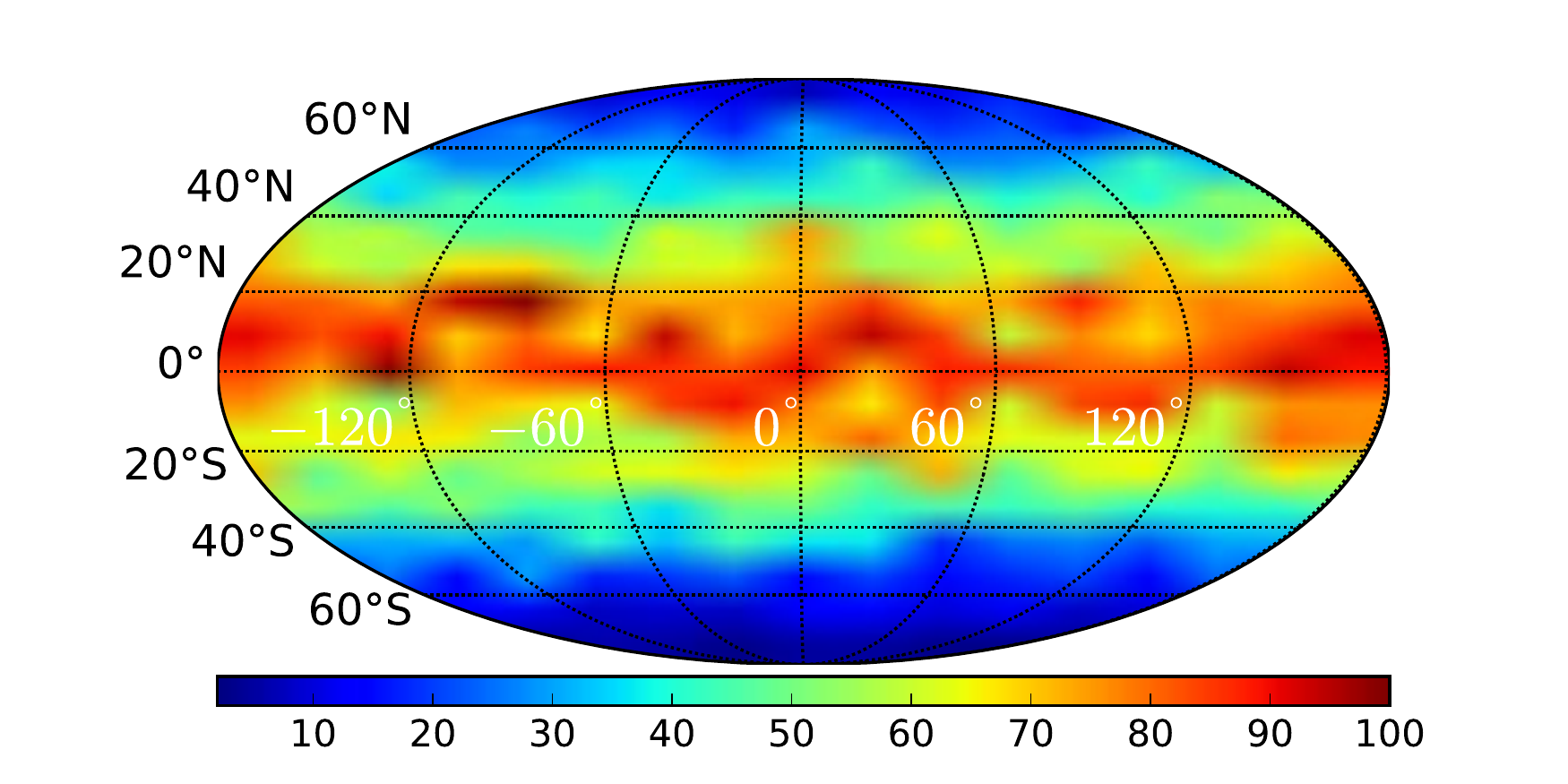}
    \includegraphics[width=0.49\linewidth]{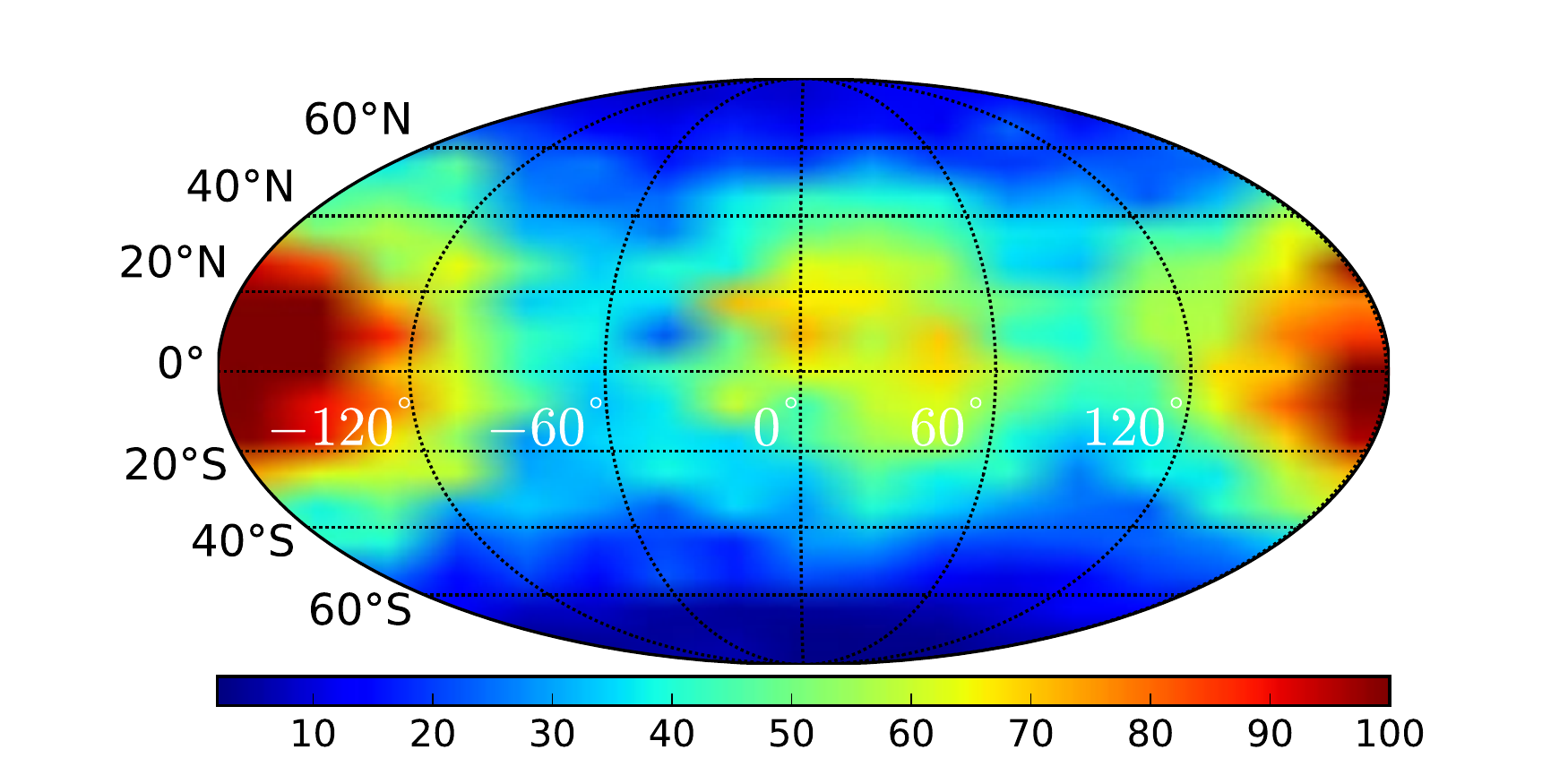}
    \caption{Angular distribution of hypervelocity stars for mass ratios $q=1$ (top row), $q=1/3$ (middle row), $q=1/9$ (bottom row) and initial eccentricities $e=0.01$ (left column) or $e=0.9$ (right column). The majority of stars are directed close to the orbital plane of the binary, and in the case of an eccentric binary, along its semimajor axis. }
    \label{fig:angdist}
\end{figure}

The stars involved in the three-body interaction events involving a hard binary are accelerated to high velocities \citep{YuTremaine2003,Sesana2006,Darbha2019}, occasionally exceeding the escape velocity from the galaxy centre ($\sim10^3$~km/s) by large factors. These stars would quickly leave the host galaxy and travel in the intergalactic space with large enough speeds that their proper motion could be detectable even at large distances \citep{GuillochonLoeb2015}.

Figure \ref{fig:veldist} shows the velocity distribution function of stars after 1~Gyr of evolution, illustrating the emergence of an unbound population of stars reaching velocities up to $10^4$~km/s. Their total number is proportional to the mass of the binary, with a weak dependence on the mass ratio, as explained in the next section. The shape of the velocity profile is roughly the same in all cases and is close to exponential, with a characteristic scale $1000-1500$~km/s comparable to the orbital speed of the binary during most of its evolution. In the case of eccentric binary, a small fraction of stars happens to be ejected while the binary is at the pericentre of its orbit, acquiring much higher speeds (dubbed `semi-relativistic stars' by \citealt{GuillochonLoeb2015}). 

Figure \ref{fig:angdist} illustrates the angular distribution of hypervelocity stars with $v>1300$~ km/s, in an equal-area projection. The binary orbit lies in the equatorial plane, and the semimajor axis is directed at longitude $0^\circ$. Most of the stars are ejected in directions close to the orbital plane of the binary, in agreement with \citet{ZierBiermann2001} and \citet{Sesana2006}. The azimuthal distribution is naturally uniform in the case of small eccentricity, but becomes more biased towards the direction of the binary semimajor axis in the high-eccentricity case (peaks at longitudes $0^\circ$ and $180^\circ$). This appears to disagree with the statement made in \citet{Sesana2006}; however, they do not plot the angular distribution itself, only quote its moments (mean azimuthal angle and its dispersion), and these numbers are in good agreement with our results. Recently \citet{Rasskazov2018} independently found an enhanced proportion of ejected stars near the orbital plane and along the direction of the binary semimajor axis (their Figure~4).

To verify our calculations, we also performed a suite of scattering experiments with isolated binary SMBH interacting with an isotropically distributed population of incoming stars; the angular distribution of ejected stars was similar to the one obtained in the fully self-consistent simulations of binary evolution. In our approach, the orientation of the binary semimajor axis is assumed to be fixed, but in reality it precesses in the orbital plane (even when the plane itself is largely conserved), hence the azimuthal distribution is averaged over time to a nearly uniform one.
Finally, we note that the angular distribution is nearly independent of the radius at which it is measured, indicating that the triaxial galactic potential is unable to significantly deflect these high-velocity stars.

\subsection{Evolution of the stellar distribution}  \label{sec:massdef}

\begin{figure}
    \includegraphics{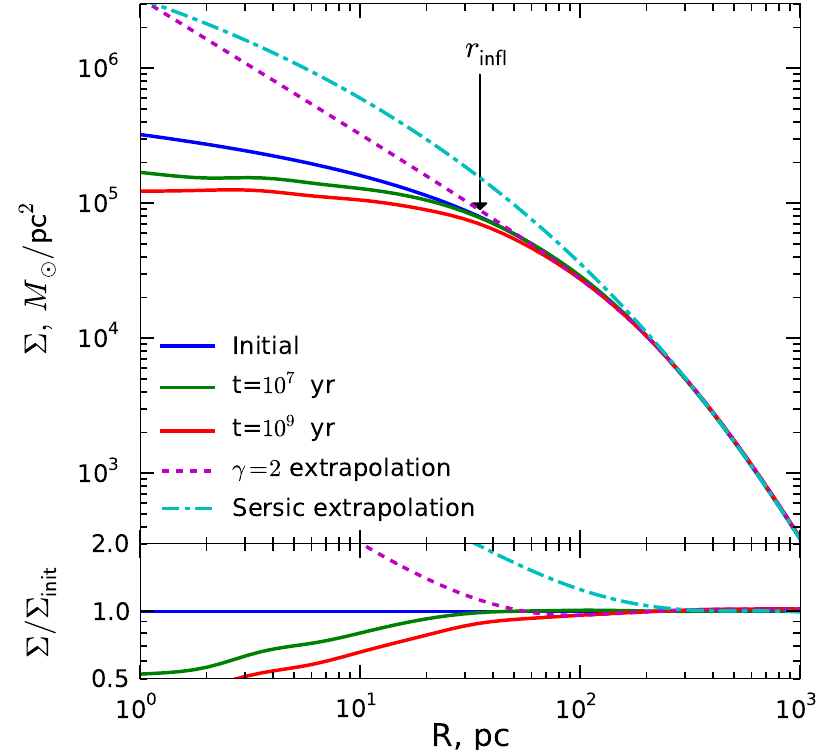}
    \caption{
    Surface density profiles of triaxial models with $M_\bullet=10^8$, $q=1/9$, at different stages of evolution: initial (Dehnen $\gamma=1$ profile, blue), shortly after the formation of a hard binary (semimajor axis $a\simeq 1$~pc, green), and after $10^9$~yr of evolution ($a\simeq 0.03$~pc, red).For the latter profile, we additionally show two models which aim at reconstructing the `original' (non-depleted) density under various assumptions. Dashed magenta line shows the density profile continued as $\rho \propto r^{-2}$ inwards from $R\simeq 100$~pc (the radius at which the logarithmic slope of the 3d density profile equals $-2$, as in \citealt{Milos2002}). Dot-dashed cyan line shows the S\'ersic profile with shape parameter $n\simeq 4.5$, which is extrapolated from the outer part of the actual density profile. More precisely, a core--S\'ersic model is fitted to the actual profile (it matches the red curve reasonably well), and then the core is undone, as in \citet{DulloGraham2012,DulloGraham2014} and \citet{Rusli2013}. Both these extrapolated reconstructions bear little resemblance to the actual initial profile, and they yield overestimated mass deficits ($\sim 4\,M_\bullet$ for $\gamma=2$ and $\sim 18\,M_\bullet$ for the S\'ersic models), while the actual missing mass is $\sim 0.4\,M_\bullet$ at $t=10^7$~yr and $\sim 2\,M_\bullet$ at $t=10^9$~yr.}
    \label{fig:mass_deficit}
\end{figure}

\begin{figure}
    \includegraphics[width=0.99\linewidth]{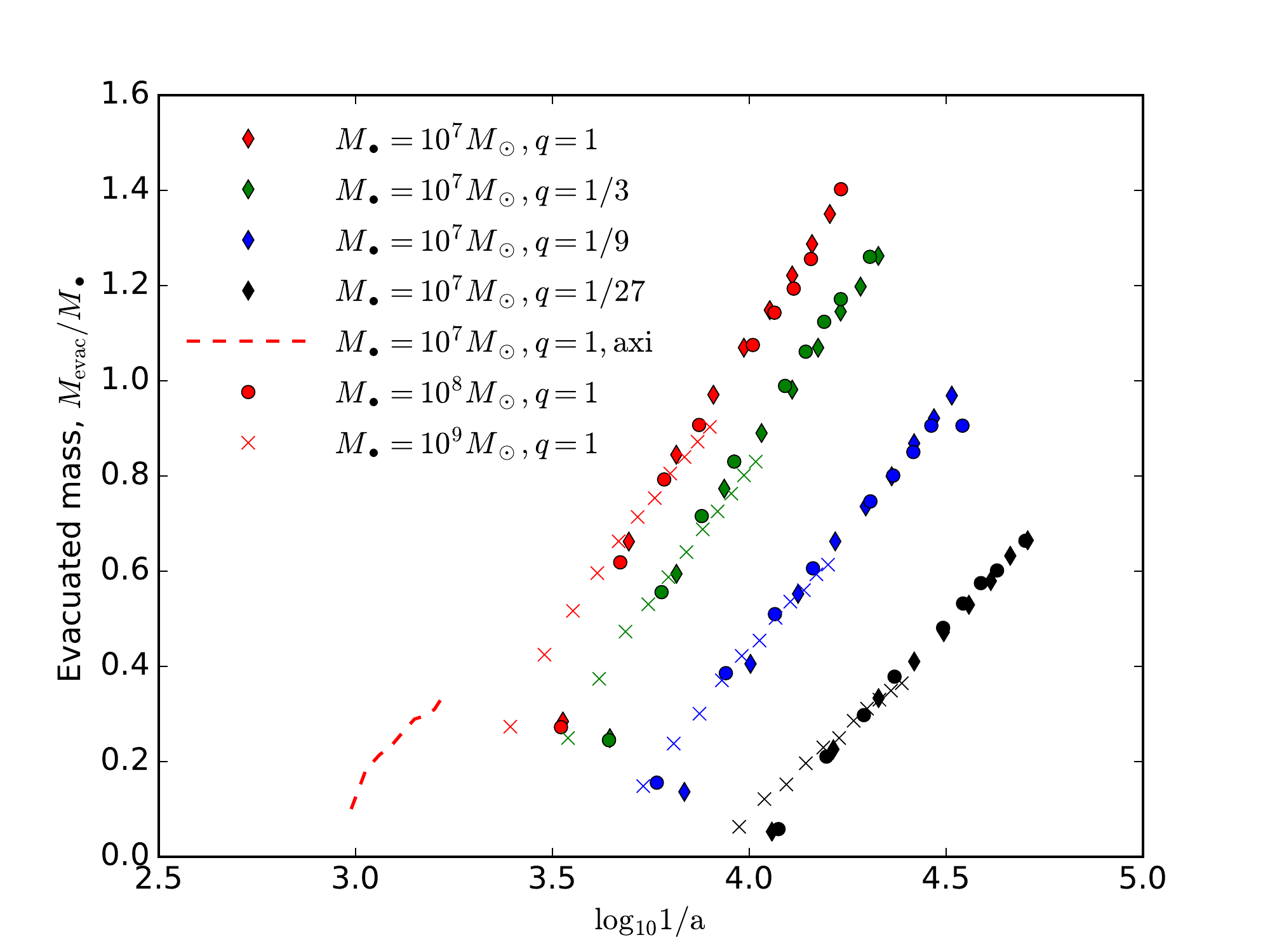}
    \caption{Evolution of evacuated mass as a function of inverse binary semimajor axis. Solid lines correspond to the total mass of the binary $M_\bullet = 10^7\, M_\odot$, circles -- $10^8\, M_\odot$, crosses -- $10^9\, M_\odot$; different colours correspond to mass ratio $q=1$ (red), $1/3$ (green), $1/9$ (blue), and $1/27$ (black).  Total evacuated mass at the end of simulation (3 Gyrs of evolution) in triaxial case scales with q as $M_{\rm evac} \approx M_\bullet\,q^{1/4}$. Dashed line represents the simulation with $M_\bullet=10^7\, M_\odot$ and $q=1$ in an axisymmetric stellar potential, in which case the evolution of semimajor axis does not progress too far; the case of a spherical potential is not shown, as the evolution is negligible. }
\label{fig:evac}
\end{figure}

Stars that have been scattered by the binary are ejected from the galaxy, leaving a `mass deficit' in its central part. The depleted density profile in the centre has long been considered a signature of the binary being present in the galaxy currently or at some time in the past \citep[e.g.][]{Milos2002, Volonteri2003, Graham2004, Merritt2006}. While we confirm that some depletion takes place, it does not appear to have a pronounced and unambiguously detectable effect on the density profile.

The central mass deficit is usually defined as the difference between the actual surface density profile $\Sigma(R)$ and the reconstructed initial profile before the binary evolution had taken place. The trouble is, of course, that this initial profile is not known, and the inference about missing mass strongly depends on the assumptions about the original un-depleted profile. 
For instance, \citet{Milos2002} deprojected $\Sigma(R)$ of a sample of galaxies to obtain the intrinsic density profiles $\rho(r)$, then for each galaxy located the radius $r_2$ where the negative logarithmic slope $\gamma\equiv -\mathrm{d}\rho/\mathrm{d}r$ attains the value 2, and extrapolated the initial profile inwards as $\rho_\mathrm{init} = \rho(r_2)\,(r/r_2)^{-2}$. This assumption of an initial steep cusp is adequate for low-mass galaxies, but if the actual density was shallower, this procedure will overestimate the amount of missing mass; indeed their inferred mass deficits change by a factor of few when the inner slope is reduced to 3/2. Moreover, there is no fundamental reason why the initial density profile would follow a power law with any particular slope.

\citet{Graham2004} assumes that the un-depleted galaxy follows the S\'ersic density profile, and the deviations from it may be parametrized by a core-S\'ersic model (Equation~5 in \citealt{Graham2003}). Thus the surface density profile is fit by the core-S\'ersic model, and its structural parameters (S\'ersic index $n$ and effective radius $R_e$) are assumed to characterize the un-depleted profile. He argues that the S\'ersic model provides a smoothly varying logarithmic slope, and is hence able to describe $\Sigma(R)$ of some galaxies without the need to introduce a core (Figure~3b in \citealt{Graham2004}). However, the (non-cored) S\'ersic model implies a rigid link between the density slopes in the central parts and in the outskirts of the galaxy: the larger is the parameter $n$, the steeper is the central profile and the shallower is the outer profile. In the core-S\'ersic model, this relation is removed, and the index $n$ is mainly determined by the outer profile, while the size of the core and the steepness of the transition are determined by the density profile in central parts. So far there is nothing wrong about such fitting procedure; however, when the mass deficit is inferred as the integrated difference between the cored and non-cored profiles with the same $n$ and $R_e$, it depends strongly on the parameter $n$, i.e., on the properties of the galaxy in the outer parts, which have no physical relation to the region around the binary. Figure~1 in \citet{HopkinsHernquist2010} illustrates this point: if a galaxy with an initial S\'ersic profile experiences a merger (without any secondary SMBH), the accreted material in the outer parts makes the density profile shallower, and hence increases the S\'ersic index.  Even though there is no change in the actual density profile in the inner part, the core-S\'ersic model would imply a significant mass deficit w.r.t.\ a steeper cusp resulting from the extrapolation of the outer profile inward.

Figure~\ref{fig:mass_deficit} shows the evolution of surface density profile in a simulation of a triaxial galaxy with $M_\bullet=10^8$ and $q=1/9$. After the formation of a hard binary (semimajor axis $a\simeq a_\mathrm{h}$, Equation~\ref{eq:ah}), the density is somewhat reduced within $R\lesssim r_\mathrm{infl}$), and after 1~Gyr the semimajor axis shrinks by another factor of 30, and the density is further depleted within a region a few times larger. However, there is no dramatic difference between density profiles at any stage of evolution. The reason is that the stars interacting with the binary are supplied from relatively large radii by centrophilic orbits, and hence their depletion has a very weak effect ($\lesssim 10\%$) on the density profile at these radii. For comparison, we show the reconstructed `un-depleted' profiles using either the $\gamma=2$ inward extrapolation or the core-S\'ersic fit: both approaches grossly misrepresent the actual initial profile. The inferred mass deficit is actually not too far off when using the $\gamma=2$ extrapolation, because an overestimate of the initial density at $R\lesssim r_\mathrm{infl}$ is partly compensated by a slight underestimate at $R\gtrsim r_\mathrm{infl}$, while the core-S\'ersic fit implies a $10\times$ larger mass deficit than the true value. Unsurprisingly, fitting a core-S\'ersic model to the initial profile also implies a presence of a core with roughly the same amount of missing mass. Moreover, varying the strength of the transition between the core and the outer S\'ersic profile leads to a few-fold change in the estimated mass deficit, even though the variation of the density between the fitted profiles is only of order 1\%.

The bottom line is that estimating the mass evacuated from the central part of the galaxy based on its present-day density profile is nearly impossible. Not only the original density profile is unknowable, but also the mass deficit is highly sensitive to the behaviour of the fitted profile near the transition radius. Even though this could not be done for any individual galaxy because of large scatter \citep[e.g., Table 4 in][]{Rusli2013}, nevertheless the general trends could be explored by analyzing a large ensemble of galaxies. \citet{HopkinsHernquist2010} assess the typical amount of missing mass by comparing the averaged profiles of cored and non-cored galaxies, and infer it to be comparable to $M_\bullet$ (note however that $M_\bullet$ in their list of galaxies is not measured directly but estimated from existing relations with other galaxy structural parameters).

On the theory side, \citet{Merritt2006} explored the evolution of unequal-mass binaries $(1/40 \le q \le 1/2$) in spherical galaxies with initial density profiles following the Dehnen model with $\gamma=1/2, 1, 3/2$, using direct-summation $N$-body simulations with $N\le 2\times10^5$. In a spherical system, the binary evolution is expected to stall at a semimajor axis $a_\mathrm{stall}$ that is only moderately smaller than $a_\mathrm{h}$. He computes the mass deficit as the difference between the (known) initial density profile and the one at the moment when $a=a_\mathrm{stall}$. He finds that the mass deficit roughly equals half the total mass of the binary, and weakly depends on the mass ratio: $M_\mathrm{def} \simeq 0.7\, M_\bullet \,q^{0.2}$. 
If the binary continues to shrink due to the interaction with stars (and not because of other mechanisms such as gas drag or gravitational waves), the evacuated mass is expected to increase, roughly proportional to $J\,M_\bullet\,\ln[a_\mathrm{init}/a(t)]$, where $J\simeq 0.5$ is the dimensionless mass ejection rate coefficient \citep{Quinlan1996, Sesana2006}.

Figure~\ref{fig:evac} shows the evolution of evacuated mass $M_\mathrm{evac}$ in our simulations of triaxial galaxies with various values of $M_\bullet$ and $q$, confirming a nearly-linear growth of $M_\mathrm{evac}$ with $\ln 1/a$. In this plot, $M_\mathrm{evac}$ is computed as the total mass of particles with positive energy (ejected from the entire galaxy). This is likely an underestimate of the mass deficit, because not all stars scattered by the binary acquire velocities greater than the escape speed (although most of them do at later stages, when $a \ll a_\mathrm{h}$). Another approach is to compute the difference between the initial density profile and the one at the given time (e.g., \citealt{Khan2012}, Section~4), which produces up to $\sim2\times$ higher estimate of the ejected mass, in agreement with a similar analysis of \citet[their Equation~41]{MilosMerritt2001}. We note in passing that in triaxial systems, stars ejected with less-than-escape speed are unlikely to return back to the centre and experience a secondary slingshot, being deflected by the large-scale torques in the potential.

\begin{figure}
    \includegraphics{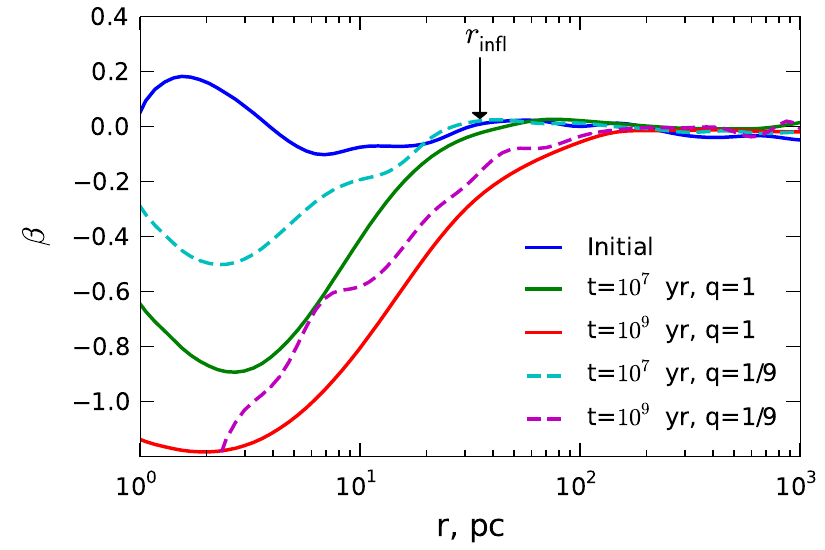}
    \caption{Radial profiles of velocity anisotropy coefficient $\beta$ for triaxial models with $M_\bullet=10^8\,M_\odot$ and $q=1$ (solid) or $q=1/9$ (dashed lines). The initial model (blue) is close to isotropic, but as the binary ejects stars preferentially from radial orbits, the velocity distribution becomes more tangentially anisotropic in the central region in the course of evolution.}
\label{fig:anisotropy}
\end{figure}

\begin{figure}
    \includegraphics{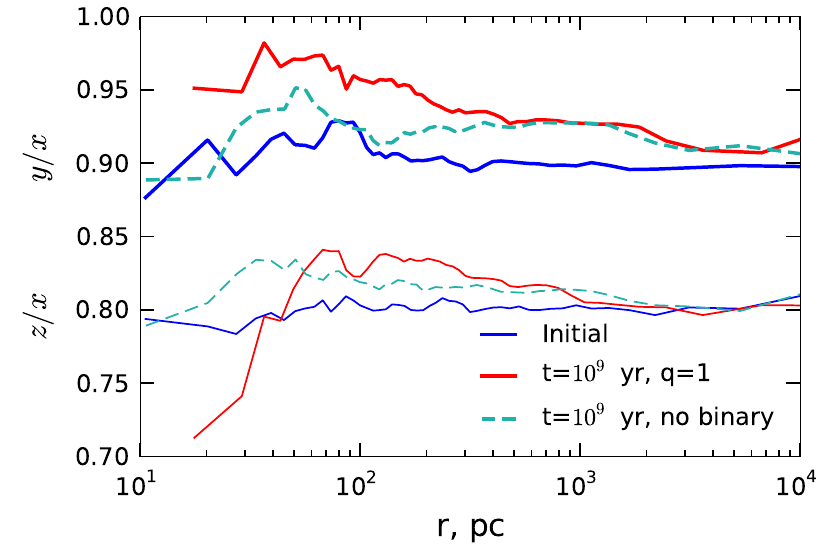}
    \caption{Radial profiles of the axis ratios for triaxial models with $M_\bullet=10^8\,M_\odot$ and $q=1$ (they are almost insensitive to $q$). The initial model (blue line) had a constant axis ratio $x:y:z = 1:0.9:0.8$, and after 1~Gyr of evolution (red line) still remains substantially triaxial despite the presence of the binary. For comparison, we show the shape of a model with a single black hole after 1~Gyr of evolution (dashed cyan line), which also did become somewhat more axisymmetric.}
\label{fig:shape}
\end{figure}

Since the stars ejected after interacting with the binary arrive from nearly-radial orbits, their disappearance creates a tangential velocity anisotropy in the central parts of the galaxy. Figure~\ref{fig:anisotropy} shows the radial profile of the anisotropy coefficient, $\beta \equiv 1 - \sigma_t^2/(2\sigma_r^2)$, where $\sigma_r$ and $\sigma_t$ are the radial and tangential velocity dispersions. The initial models were constructed to be nearly-isotropic, but the depletion of radial orbits leaves a pronounced trace in the anisotropy profile in the same range of radii where the density profile is modified (up to a few influence radii). The magnitude of tangential anisotropy is in qualitative agreement with the simulations of \citet[their Figure~7]{QuinlanHernquist1997}, \citet[their Figure~16]{MilosMerritt2001} and \citet[their Figure~7]{Rantala2018}.
We stress that this velocity anisotropy coefficient is not directly observable, but has to be inferred from dynamical models (at least when its value is not assumed or kept constant in the modelling procedure, as often done in the Jeans approach). In fact, it is hard to find any direct signature of core depletion in the observable values -- the most apparent one is the lack of a central spike in the line-of-sight velocity dispersion profile in tangentially-anisotropic systems. Using the Schwarzschild method, \citet{Thomas2014} did find an indication for tangential anisotropy in central parts of several galaxies with cored density profiles, which may be regarded as a more substantial evidence for the scouring effect of the binary than the core alone.

Finally, given that the continued shrinking of the binary depends crucially on the existence of centrophilic orbits in a triaxial potential, it is worthwhile to check whether it remains triaxial over the entire evolution. Figure~\ref{fig:shape} plots the radial profile of the axis ratio ($y/x,\;z/x$) at the initial moment and after 1~Gyr of evolution (for a $M_\bullet=10^8\,M_\odot$, $q=1$ model, but the results are similar in other cases). It demonstrates that the system evolves towards a more axisymmetric shape ($y/x \sim 0.95$ in the central parts, higher than the initial value 0.9), but remains sufficiently triaxial over its lifetime.

\section{Discussion and conclusions}  \label{sec:summary}

In this paper, we study the long-term co-evolution of binary SMBHs and galactic nuclei, using a large suite of stellar-dynamical simulations conducted with the Monte Carlo code \textsc{Raga}.
We explore the influence of several parameters of the binary (total mass, initial eccentricity, mass ratio, inclusion or neglect of stellar captures and GW losses) and the stellar cluster (shape and rotation). We start the simulations from equilibrium initial models for the stellar distribution, at the moment of formation of a hard binary, and conduct them for a duration of time sufficient for the binary to reach the GW-dominated regime (up to a few Gyr).
Our main conclusions are the following:

\begin{itemize}
\item The stellar-dynamical hardening rate of the binary is almost independent of the initial eccentricity, and is nearly the same whether we allow the stars to be captured or not. Of course, if one takes the GW losses into account, a highly eccentric binary would reach coalescence sooner.

\item The hardening rate depends weakly on the binary mass ratio $q$, but much more strongly on the geometry of the stellar cluster (spherical, axisymmetric or triaxial). Only in the latter case the binary is able to reach the GW-dominated regime in a time much shorter than the Hubble time, regardless of its initial parameters, in agreement with \citet{Vasiliev2015b}, \citet{Gualandris2017}. The hardening rate is significantly lower (by a factor $3-10$) than the so-called `full loss-cone rate' due to the depletion of centrophilic orbits, however, their reservoir is large enough to sustain the long-term evolution of the binary. The $q$-dependence of the hardening rate seen in Figure~\ref{fig:sma} arises primarily from the fact that a smaller secondary SMBH leads to a less dramatic destruction of the stellar cusp at the early stages of evolution. However, the coalescence time (Equation~\ref{eq:tcoal}) is nearly independent of $q$.

\item The eccentricity evolution depends on its initial value and on the rotation and geometry of the stellar cluster. We confirm previous results regarding the strong tendency of eccentricity to increase or decrease in counterrotating or corotating stellar clusters, correspondingly \citep[e.g.,][]{Sesana2011, HolleyKhan2015}. However, in triaxial clusters this growth or decay saturates more quickly, because the principal supply source of stars into the loss cone comes from centrophilic orbits, which may not have net rotation. We also find that the eccentricity evolution in non-rotating clusters is rather stochastic, which has been observed in many previous studies \citep[e.g.][]{Wang2014, Khan2018}, but on average it tends to increase if started from a sufficiently high initial value ($\gtrsim 0.3-0.5$) or in systems with more unequal mass ratios.

\item The capture rates of stars by binary SMBH are higher than in otherwise equivalent systems with a single SMBH at the early stage of its evolution, in agreement with \citet{Chen2011}, \citet{Li2015}, \citet{Darbha2018}. However, subsequently they drop precipitously, and in the spherical and axisymmetric cases remain well below the rates expected in single-SMBH systems, again confirming the earlier analysis of \citet{Chen2008}. The decrease of capture rates parallels that of the binary hardening rates, since both are driven by the depletion of loss-cone orbits. In triaxial systems this depletion is less prominent, and the capture rates remain only slightly lower than or comparable to those in single-SMBH systems throughout the entire evolution. The capture rates do not appreciably depend on eccentricity.

\item Most of the stars that interacted with the binary at later stages of evolution are ejected from the galaxy, forming a population of hypervelocity stars with an approximately exponential distribution in speed (with a characteristic scale $\gtrsim 10^3$~km/s). In the case of eccentric binaries, the tail of this distribution extends to higher velocities, exceeding $10^4$~km/s. Their angular distribution is concentrated near the orbital plane of the binary.

\item The total mass of ejected stars is comparable to the mass of the binary, with a weak dependence on the mass ratio ($\propto q^{1/4}$), in agreement with \citet{Merritt2006} or \citet{Khan2012}. At the early stage of evolution, the initially cuspy density profile is destroyed \citep[cf.][]{MilosMerritt2001}; however, at later stages the stars entering the loss cone arrive from larger distances, and the density profile is depleted only very slightly there.
We argue that the impact of the binary on the density profile (known as core depletion or mass deficit) is very hard to assess from observations, as the difference between the (unknown) initial profile and the one modified by the binary is smaller than the uncertainties in estimating the initial profile. Most importantly, any attempt to infer the mass deficit from fitting a particular parametric model to the observed density profile cannot be accurate even to within an order of magnitude.
\end{itemize}

Our focus in this study was on the long-term evolution of stellar systems that occurs quasi-statically (the characteristic timescales for changes in the binary orbit or the stellar distribution are much longer than the dynamical time). The secular Monte Carlo approach is well-suited for such problems, although it has a number of limitations, which we believe to be insignificant for most of our conclusions:

\begin{itemize}

\item We consider only stellar-dynamical processes, neglecting any gas physics, thus our analysis is applicable only to binary SMBH formed in dry mergers.

\item We assume that the evolution proceeds through a sequence of nearly steady-state configurations. Thus we are not able to adequately simulate the very early stage when the binary just forms, or consider highly dynamic situations such as an infall of a globular cluster \citep{ArcaSedda2017, Bortolas2018a}, a third SMBH \citep{Iwasawa2006, HoffmanLoeb2007, Ryu2018}, or massive perturbers \citep{PeretsAlexander2008}. The momentarily elevated capture rates at the early stage of binary evolution \citep{Chen2011, Li2017} are also observed in our simulations, but cannot be quantified robustly. The duration of this early stage is short (a few Myr) compared to the overall lifetime of a binary ($\sim 1$~Gyr).

\item We also assume that the binary orbit is aligned in a specific plane ($x-y$) and does not change its inclination throughout the simulation, which is reasonable to expect especially in non-spherical stellar potentials \citep{CuiYu2014}. A detailed Fokker--Planck study of diffusive changes in the binary orbital parameters (eccentricity and inclination) has been conducted by \citet{RasskazovMerritt2017}, who showed that the inclination and eccentricity evolve independently, and the expected change in the inclination angle over the lifetime of the binary is moderate. 

\item In our approach, the centre-of-mass of the binary is pinned down at the origin. It was suggested by \citet{Chatterjee2003} that the Brownian motion of the binary may enhance the hardening rate. However, analytical estimates by \citet{MilosMerritt2003} and N-body simulations by \citet{Bortolas2016} demonstrate that the stochastic wandering of the binary SMBH centre-of-mass has only a minor effect ($\lesssim 10\%$) on the hardening rates. On the other hand, \citet{HolleyKhan2015} and \citet{Mirza2017} found that in stellar clusters with prograde rotation, the binary centre-of-mass settles onto a finite-size orbit around the galaxy centre; however, this global motion did not significantly affect the hardening rate, although the duration of their simulations is much shorter than considered in the present paper. 

\item We do not consider the role of binary stars, which may influence the dynamical evolution of the single/binary SMBHs in various ways. \citet{Wang2018} explored various effects in four-body interactions between binary stars and binary SMBH, showing that these may lead to elevated rates of tidal disruptions or ejection of hypervelocity stars, possibly providing another observational signature of a binary SMBH.

\item The Monte Carlo approach to two-body relaxation uses the diffusion coefficients computed in a spherical isotropic background, while the systems under study are in general non-spherical and anisotropic (especially in cases of strong rotation). Nevertheless, the effect of bulk rotation and the global shape of the potential is adequately taken into account at the level of stellar orbits (in the collisionless sense). Since the evolution of galactic nuclei in the mass regime considered in this paper ($M_\bullet\ge 10^7\,M_\odot$) is mainly caused by collisionless processes, this slight inconsistency in the treatment of collisional relaxation should not have any impact on the results.

\end{itemize}

The observational evidence of binary SMBHs is scarce, although this will hopefully change when the planned space-based GW observatory \textsc{Lisa} \citep{eLISA} becomes operational. In this paper, we reaffirmed the expectation that the lifetime of SMBH binaries is much shorter than the Hubble time in realistic conditions, and explored their effects on the stellar distribution in the galactic nuclei (rates of stellar captures or tidal disruptions, ejection of hypervelocity stars and erosion of density cusps). Unfortunately, none of these effects constitute an observable smoking-gun evidence for the existence of a binary SMBH in any given galaxy.

\textbf{Acknowledgements}\quad
This work has made use of the MIPT-60 cluster, hosted by Moscow Institute of Physics and Technology. KL thanks S.V. Ermakov for helpful discussions.

\label{lastpage}

\end{document}